%% file: main.tex
\documentclass[
]{ceurart}

\usepackage{listings}
\usepackage{comment}
\usepackage{adjustbox}
\usepackage{tabularx}

\usepackage{tikz}
\usepackage{fontawesome5} 
\usetikzlibrary{arrows.meta, positioning, fit, backgrounds, calc}

\begin{document}

\copyrightyear{2025}
\copyrightclause{Copyright for this paper by its authors.
  Use permitted under Creative Commons License Attribution 4.0
  International (CC BY 4.0).}

\conference{Joint Proceedings of the ACM IUI Workshops 2026, March 23-26, 2026, Paphos, Cyprus}

\title{From Prompt to Product: A Human-Centered Benchmark of Agentic App Generation Systems}

\author[1]{Marcos Ortiz}
\fnmark[1]
\address[1]{Quome Inc., West Hollywood, California}

\author[1]{Justin Hill}
\fnmark[1]

\author[1]{Collin Overbay}

\author[1]{Ingrida Semenec}

\author[1]{Frederic Sauve-Hoover}

\author[1]{Jim Schwoebel}

\author[2, 1]{Joel Shor}[
email=joel.shor@move37labs.ai,
url=https://move37labs.ai/,
]
\cormark[1]
\address[2]{Move37 Labs, Somerville, MA, USA}

\fntext[1]{These authors contributed equally.}
\cortext[1]{Corresponding author.}

\input{sections/0_abstract}

\begin{keywords}
    prompt-to-app \sep
    agents \sep
    agentic AI \sep
    benchmark \sep
    generative software \sep
    human-centered \sep
    low-code \sep
    no-code
\end{keywords}

\maketitle

\input{sections/1_introduction}

\input{sections/2_background}

\input{sections/3_methodology}

\input{sections/4_results}

\input{sections/5_discussion}

\input{sections/6_conclusion}

\begin{acknowledgments}
  We are grateful to Hong Joo Ryoo for insightful discussions that informed the design of our automation pipeline for the Firebase Studio system. Thanks to Quome Inc. for funding this study.
\end{acknowledgments}

\section*{Declaration on Generative AI}
 During the preparation of this work, the author(s) used Gemini-3, and GPT-5.2, in order to: draft the introduction, improve writing style, grammar and spell checking, formatting assistance, and peer review simulation. Sonnet 4.5 and Composer-1 were used to assist in writing or improving code for automation of app generation, and to assist in creating and formatting charts and tables. After using these tool(s)/service(s), the author(s) reviewed and edited the content as needed and take(s) full responsibility for the publication’s content. 

\bibliography{bibliography}

\appendix
\section{Appendices}
\input{sections/7_appendix}

\end{document}

%% file: sections/0_abstract.tex
\begin{abstract}
Agentic AI systems capable of generating full-stack web applications from natural language prompts (``prompt-to-app'') represent a significant shift in software development. However, evaluating these systems remains challenging, as visual polish, functional correctness, and user trust are often misaligned. As a result, it is unclear how existing prompt-to-app tools compare under realistic, human-centered evaluation criteria.
In this paper, we introduce a human-centered benchmark for evaluating prompt-to-app systems and conduct a large-scale comparative study of three widely used platforms: Replit, Bolt, and Firebase Studio. Using a diverse set of 96 prompts spanning common web application tasks, we generate 288 unique application artifacts.
We evaluate these systems through a large-scale human-rater study involving 205 participants and 1{,}071 quality-filtered pairwise comparisons, assessing task-based ease of use, visual appeal, perceived completeness, and user trust. Our results show that these systems are not interchangeable: Firebase Studio consistently outperforms competing platforms across all human-evaluated dimensions, achieving the highest win rates for ease of use, trust, visual appeal, and visual appropriateness. Bolt performs competitively on visual appeal but trails Firebase on usability and trust, while Replit underperforms relative to both across most metrics.
These findings highlight a persistent gap between visual polish and functional reliability in prompt-to-app systems and demonstrate the necessity of interactive, task-based evaluation. We release our benchmark framework, prompt set, and generated artifacts to support reproducible evaluation and future research in agentic application generation.
\end{abstract}

%% file: sections/1_introduction.tex
\section{Introduction}
\label{sec:introduction}

The advent of large language models has triggered a rapid evolution in software development tools, moving from simple code completion to sophisticated, agentic systems. A new frontier in this space is the ``prompt-to-app'' agent, which aims to generate entire full-stack, deployable web applications from a single natural language description. These systems promise to dramatically lower the barrier to software creation, enabling ``citizen developers'' and accelerating prototyping for experienced engineers.
However, this promise is tempered by a significant lack of rigorous evaluation. How do we know if these generated applications are good? Are they functional? Do they produce results that real users can interact with?

Existing benchmarks for code generation typically focus on small, function-level snippets or algorithmic challenges. These are insufficient for evaluating prompt-to-app agents, which must correctly orchestrate complex interactions between a frontend, a backend, and a database, all while delivering a coherent user experience.

To address this gap, we present the first large-scale, human-centered benchmark of prompt-to-app generation agents. This addresses the need for new methodologies for user-centered evaluation and real-world usefulness of agentic systems, moving beyond static benchmarks to capture the holistic developer experience.

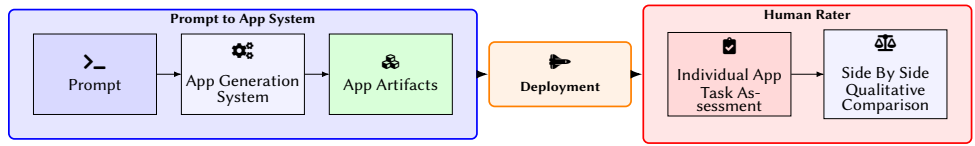
\begin{figure}[h]
    \label{fig: benchmark flow chart (prompt, deploy, rate)}
    \centering
    \begin{adjustbox}{width=0.8\linewidth}
        \input{images/flow_chart}
    \end{adjustbox}
    \caption{To evaluate prompt-to-app systems, we curated a list of prompts and used each system to generate App Artifacts. Those artifacts were then deployed, when possible, as live web applications. Then, human raters were asked to evaluate individual apps, as well as make side by side comparisons between apps generated by different systems using the same prompt.}
    \label{fig:my_image}
\end{figure}

\paragraph{Our contributions are as follows:}
\begin{itemize}
    \item A novel mixed-methods evaluation framework for prompt-to-app agents, combining automated checks with task-based human assessment.
    \item A comparative analysis of three leading systems: Replit, Bolt, and Firebase Studio.
    \item A public dataset of the 288 generated application artifacts, 96 from each platform.
    \item Key findings that characterize the current state of the field:
    \begin{itemize}
        \item Isolated evaluation of these systems proves difficult, while side by side comparisons yield clear user preferences.
        \item It is possible for a single system to dominate across several metrics, indicating that the market is not yet commoditized
    \end{itemize}
\end{itemize}

The benchmark we establish provides a critical tool for developers and researchers to navigate the ``prompt-to-app'' landscape and identifies key areas for future innovation.

%% file: images/flow_chart.tex
\begin{tikzpicture}[
    >=Stealth,
    process/.style={
        rectangle, 
        draw, 
        fill=blue!5, 
        text width=2.8cm,       
        minimum height=2.0cm,   
        align=center, 
        font=\sffamily\scriptsize 
    },
    container/.style={rectangle, draw, very thick, rounded corners, inner sep=0.6cm},
    deployment_style/.style={
        rectangle, draw=orange, fill=orange!10, very thick, rounded corners, 
        minimum width=3.5cm, minimum height=1.6cm, align=center, 
        font=\sffamily\bfseries\small 
    },
    internal_flow/.style={-Latex, thin},
    main_flow/.style={-Latex, very thick}
    ]
    
    
    \node[process] (app_gen) {
        \large\faCogs\\[0.5em] 
        App Generation\\System
    };
    
    \node[process, left=0.6cm of app_gen, fill=blue!15] (prompt) {
        \large\faTerminal\\[0.5em]
        Prompt
    };
    
    \node[process, right=0.6cm of app_gen, fill=green!15] (artifact) {
        \large\faCubes\\[0.5em]
        App Artifacts
    };
    
    \draw[internal_flow] (prompt) -- (app_gen);
    \draw[internal_flow] (app_gen) -- (artifact);
    
    \begin{scope}[on background layer]
        \node[container, draw=blue, fill=blue!10, fit=(prompt) (artifact),
        label={[anchor=north, font=\bfseries]north:Prompt to App System}] (stage1) {};
    \end{scope}
    
    
    \coordinate (right_center) at ($(app_gen) + (14cm, 0)$);
    
    \node[process, fill=red!15] (task) at ($(right_center) + (-2.0, 0)$) {
        \large\faClipboardCheck\\[0.5em]
        Individual App\\Task Assessment
    };
    
    \node[process, right=0.8cm of task] (sbs) {
        \large\faBalanceScale\\[0.5em]
        Side By Side\\Qualitative Comparison
    };
    
    \draw[internal_flow] (task) -- (sbs);
    
    \begin{scope}[on background layer]
        \node[container, draw=red, fill=red!10, fit=(task) (sbs),
        label={[anchor=north, font=\bfseries]north:Human Rater}] (stage3) {};
    \end{scope}
    
    
    \node[deployment_style] (deployment) at ($(stage1.east)!0.5!(stage3.west)$) {
        \large\faSpaceShuttle\\[0.4em]
        Deployment
    };
    
    \draw[main_flow] (stage1.east) -- (deployment.west);
    \draw[main_flow] (deployment.east) -- (stage3.west);
    
\end{tikzpicture}

%% file: sections/2_background.tex
\section{Background and Related Work}
\label{sec:background}

The evaluation of generative AI has long been a challenge, particularly as outputs become more complex and open-ended. In fields like image and text generation, metrics have evolved from simple pixel-wise or n-gram-based comparisons to more sophisticated perceptual and semantic measures. Recent work has emphasized that evaluating generative AI systems becomes increasingly difficult as outputs grow more open-ended \cite{Anetal2024}. Studies have shown that traditional pixel-based or token-based similarity metrics fail to capture human-perceived quality or usability \cite{desolda2025hcai,clinical-bertscore,batool2025usability}. This has motivated a shift toward human-centered evaluation frameworks that incorporate real user judgments, interaction behavior, and trust formation.

\paragraph{Agentic AI in Software Development.}
The concept of AI agents in software development is not new, but their capability has grown exponentially. We are moving from co-pilots that assist with line-level code to agents that can manage entire projects. This includes multi-agent systems that divide tasks (for example, frontend, backend, or testing) and require complex planning and reasoning. Our work focuses on the end-product of such agentic systems, providing a black-box evaluation of their real-world utility. A parallel line of research examines agentic or multi-agent AI systems for software engineering tasks. Prior benchmarks such as HumanEval \cite{chen2021codex}, mHumanEval \cite{raihan2025mhuman}, and SWE-bench \cite{jimenez2024swebench} demonstrate the difficulty of assessing LLM-generated code even at the function or issue level. More recent studies evaluate LLM agents on multi-step or multi-requirement tasks \cite{pradas2024agentframeworks}, highlighting persistent challenges in planning, tool use, and maintaining cross-component consistency. Our work extends these lines of inquiry by evaluating the end-to-end output of such agentic systems at the level of full-stack applications.

\paragraph{Challenges in Evaluating Generative Software.}

Recent work underscores that evaluating generated software requires both technical and human-centered perspectives. Frameworks for assessing LLM-powered GUI agents stress the importance of interaction fidelity, user trust, and real-world task completion \cite{chen2025guifw}. These findings reinforce the need for benchmarks that capture usability, visual quality, and functional correctness together—dimensions that traditional software engineering benchmarks do not cover. Our work seeks to build a benchmark that measures these user-centered qualities alongside technical correctness.

%% file: sections/3_methodology.tex
\section{Benchmark Methodology \& Experimental Setup}
\label{sec:methods}

Our methodology is designed to be comprehensive, capturing both technical quality and human-perceived value. It consists of three main pillars: the systems we test, the tasks we assign, and the metrics we use to judge the results.

\paragraph{Systems Under Test (SUTs):} We selected three representative systems that cover different approaches to app generation: Replit, Bolt, and Firebase Studio. Each system is designed to generate web applications based on a user prompt. 

\paragraph{The Prompt Set:} We developed a set of 96 unique prompts designed to test a wide range of application types. Because realistic usage is essential for evaluating prompt to app systems, prompts were constructed from real world scenarios inspired by developer forums and practical software requests. To ensure systematic coverage and avoid the pitfalls of narrow evaluation sets, we adopted a multi-dimensional taxonomy approach similar to recent large-scale generative benchmarks~\citep{nucleobench}. We organized the prompts into a taxonomy defined along several axes: category (healthcare, legal, real estate, financial services, government, education), difficulty (easy, hard), specificity (specific, ambiguous), and complexity (low, high). Details on the prompt taxonomy, and examples of specific prompts, can be found in Appendix~\ref{sec:appendix_prompts}.

\paragraph{Evaluation Metrics:} To assess qualitative aspects of user experience, we employed a two-stage evaluation design using human raters. In the \textbf{Isolated stage}, raters evaluated each application independently using direct five-point scales (1=very poor, 5=very good) on \textbf{clarity} and \textbf{ease of use}. In the \textbf{Comparison stage}, raters viewed both applications side-by-side and provided comparative judgments on a five-point preference scale (1=strongly prefer App A, 3=no preference, 5=strongly prefer App B): \textbf{visual appeal} (aesthetic attractiveness), \textbf{visual appropriateness} (whether the design suits the task context), \textbf{comparative ease}, and \textbf{trust}. At both stages, we recorded whether applications successfully appeared on screen to account for technical failures.

\paragraph{Artifact Generation:} We generated a total of 288 app artifacts. To ensure consistent evaluation, all artifacts were deployed to a standardized containerized environment (Quome Cloud). See Appendix \ref{subsec:artifact_deployment_details} for details on the deployment process. System-specific conditions for artifact generation are described in Appendix~\ref{subsec:artifact_generation_details}.

\subsection{Human Rater Study Design}
\label{subsec:humanraterstudydesign}

\begin{figure}[htbp]
    \centering
    \begin{minipage}{0.3\textwidth}
        \centering
        \includegraphics[width=\textwidth]{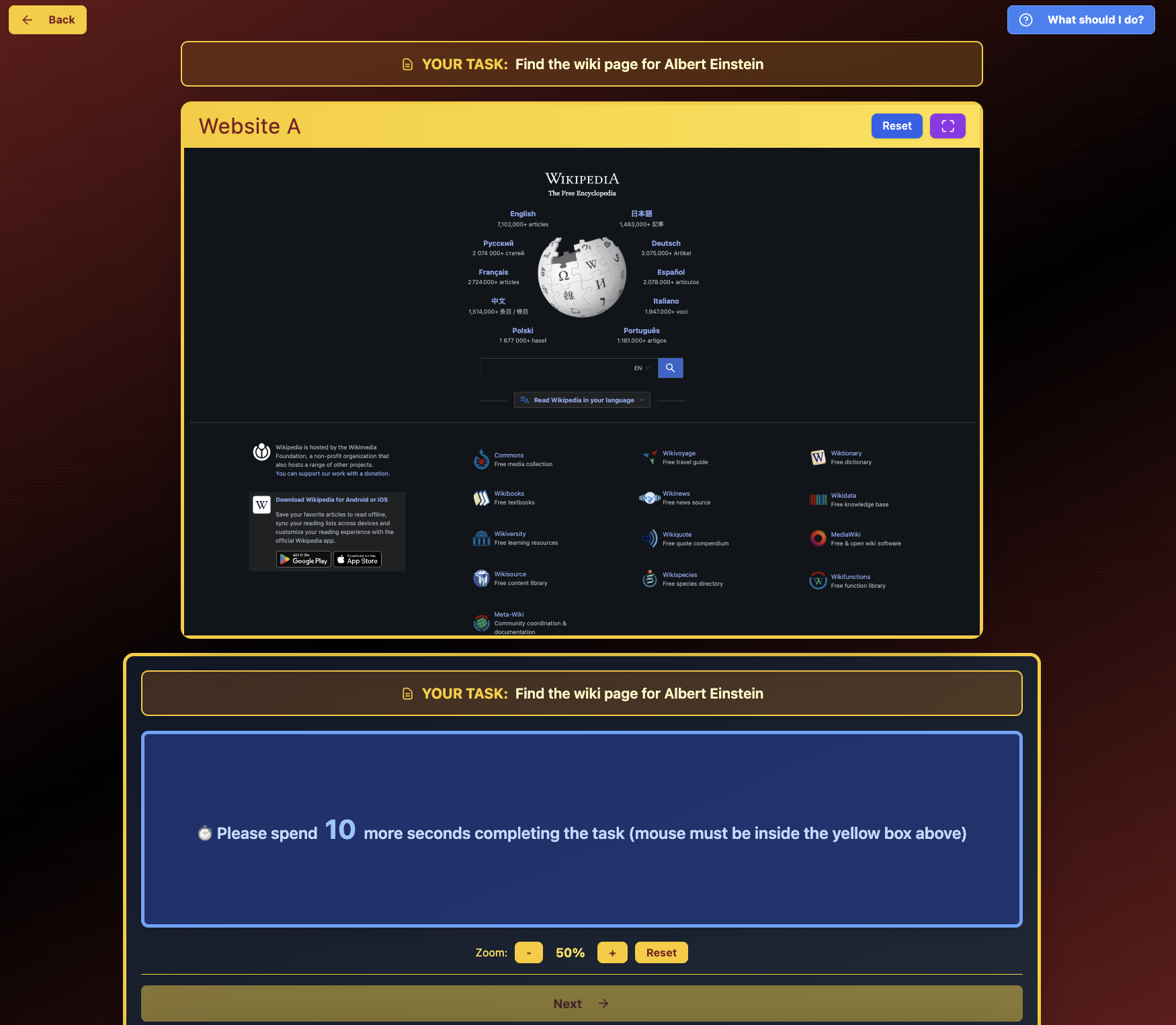}
    \end{minipage}
    \hfill 
    \begin{minipage}{0.3\textwidth}
        \centering
        \includegraphics[width=\textwidth]{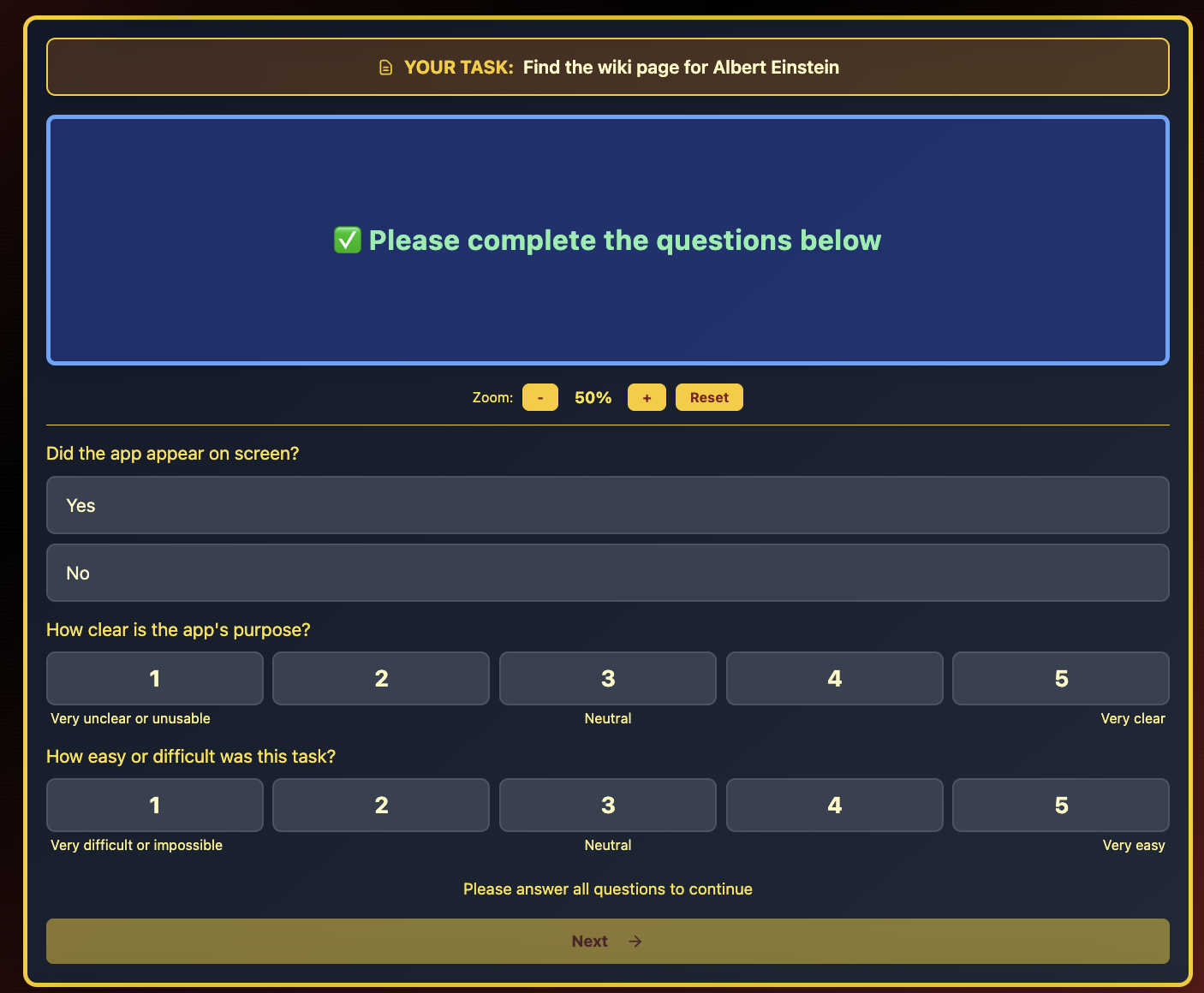}
    \end{minipage}
    \hfill 
    \begin{minipage}{0.2\textwidth}
        \centering
        \includegraphics[width=\textwidth]{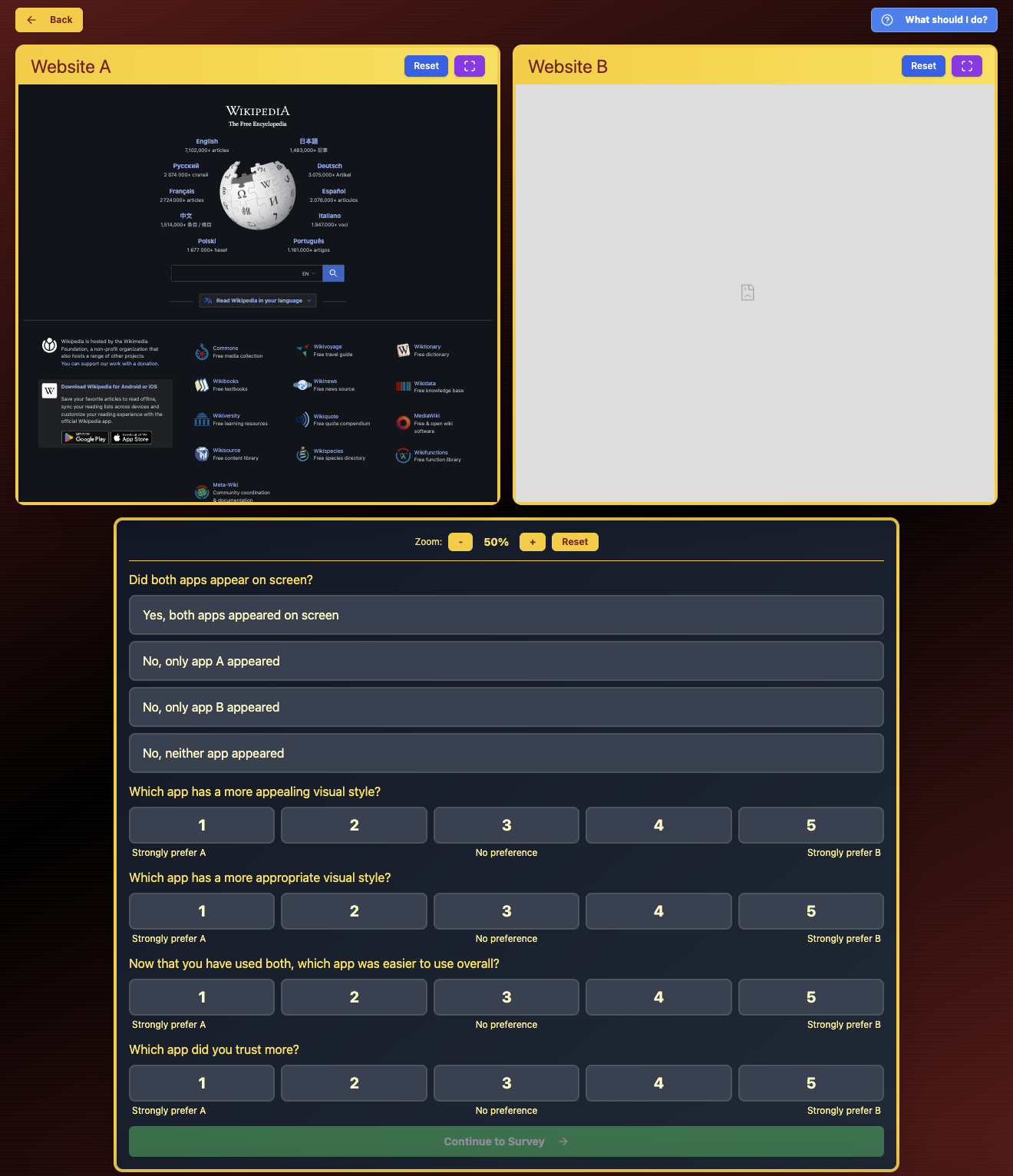}
    \end{minipage}
    
    \caption{Screenshots from the survey. \textbf{Left)} Isolated app evaluation (the example is from the tutorial). \textbf{Center)} Questions in isolated evaluation portion. \textbf{Right)} Comparison evaluation, presentation and specific questions.}
    \label{fig:three_images_side_by_side}
\end{figure}

\paragraph{Participant Recruitment.}
We recruited 282 participants through the CINT Exchange survey platform. We did not impose eligibility criteria such as technical background, in order to capture ratings from a broad range of users. The final pool was demographically diverse (Appendix~\ref{app:demographics}).

\paragraph{Refined Task Based Design.}
We developed a structured, multi stage evaluation. Each participant evaluated multiple pairs of applications through a three-phase process. First, participants interacted with App A in isolation (Figure~\ref{fig:three_images_side_by_side} left and center). Second, participants independently evaluated App B using identical measures. Finally, participants viewed both applications side-by-side and provided comparative judgments (Figure~\ref{fig:three_images_side_by_side} right). This quantitative study design, as opposed to open-ended qualitative feedback, adopts the human-centered evaluation protocols used in clinical evaluation~\citep{patient-centric}. See Appendix \ref{appendix:survey_details} for details of the survey design.

Before the main evaluation, each participant completed a short tutorial consisting of a mock comparison between two placeholder websites. This familiarized raters with the survey interface, the side by side format, and the preference scale, ensuring they understood the task before assessing generated applications.

\paragraph{Filtering for Quality Responses:} To ensure data quality, we applied strict filtering criteria: a minimum time-on-task, a cap on total amount of time to prevent fatigue, a maximum time-per-question to prevent bots and gamification, and a ``smoke-test'' involving a control question with a known answer. Participants were excluded for failing these (see Appendix~\ref{appendix:filtering} for details).

\subsection{Automated Deployment Audit}

We performed an automated audit of the deployed applications, to try and determine which had deployed successfully. The audit used a two-stage approach to determine deployment status. First, an HTTP GET request is made to detect connection errors, timeouts, and SSL errors. For URLs returning HTTP 200, a browser-based check was performed using Playwright \cite{playwright2025} to detect common visible DOM elements, or browser crashes.

\subsection{Statistical Methodology}

Our statistical analysis employed a two-tiered approach based on the stage of evaluation. For individual app metrics (appearance rate, clarity, and ease of use), we analyzed data from isolated presentation stages where participants rated each application independently. These isolated-stage ratings were included only when participants confirmed the application had appeared successfully (number of data points after filtering described in Appendix Table~\ref{tab:isolated_by_platform}). We applied the Kruskal-Wallis H test to assess overall differences across platforms, and when sufficient paired data existed, we used the Friedman test for repeated measures. For more details, see Appendix~\ref{appendix:stats}.

For comparative metrics (overall ease of use, trust, visual appeal, and visual appropriateness), we analyzed pairwise comparison data, restricting our analysis to cases where both applications in a comparison successfully appeared to participants (number of data points after filtering described in Appendix Table~\ref{tab:platform_pairs}). We used the Wilcoxon signed-rank test for hypothesis testing, and Bradley-Terry models treating the evaluation as a pairwise tournament to estimate relative platform abilities. To establish global platform rankings while controlling for confounds, we fitted Linear Mixed-Effects Models (LMM) with platform and presentation position as fixed effects and random intercepts for participants (to control for harsh vs. lenient grader bias) and prompts (to control for task difficulty). For more details, see Appendix~\ref{appendix:stats}.

%% file: sections/4_results.tex
\section{Results}
\label{sec:results}

Our experiment yielded a rich dataset of automated and human-generated results. In this section, we report high-level patterns from the automated pipeline and the human-rater study. 

We collected optional demographic information to contextualize the backgrounds of the human users who agreed to participate. Participants varied in age, gender, education level, industry background, and familiarity with AI or app building tools. The exact distributions are reported in Appendix \ref{appendix:survey_details}. These demographic fields will allow the analysis of whether preferences vary across experience levels. 

\subsection{Participant Demographics}
\label{sec:demographics}

We recruited a set of demographically diverse participants (see Appendix~\ref{app:demographics} for details), including both technical and non-technical users from industries such as healthcare and software.

\subsection{Automated Deployment Audit}

To validate that participant ratings reflect actual deployment outcomes rather than random responses, we compared automated deploy testing results with participant-reported appearance rates.

We found substantial agreement between automated testing and participant reports with an overall agreement rate of 88.2\%, as shown in Table~\ref{tab:deploy_confusion_matrix}. A chi-square test confirmed that automated and participant classifications are not independent ($\chi^2 = 133.9$, $p < 0.001$), with a strong positive correlation ($r = 0.706$, $p < 0.001$). These results provide strong evidence that participant ratings are not random and accurately reflect actual deployment success. See Appendix \ref{sec:appedix automated audit} for additional details.

\begin{table}[htbp]
    \centering
    \caption{Confusion matrix showing agreement between automated testing and participant reports. Chi-square test assesses whether automated and participant classifications are independent.}
    \label{tab:deploy_confusion_matrix}
    \begin{tabular}{llcc}
        \toprule
        & & \multicolumn{2}{c}{\textbf{Participant Report}} \\
        \cmidrule(lr){3-4}
        & & \textbf{Appeared} & \textbf{Not Appeared} \\
        \midrule
        \multirow{2}{*}{\textbf{Automated}} & Success & 197 (68.4\%) & 16 (5.6\%) \\
        & Failure & 18 (6.2\%) & 57 (19.8\%) \\
        \midrule
        \multicolumn{2}{l}{Chi-square statistic} & \multicolumn{2}{c}{133.905} \\
        \multicolumn{2}{l}{$p$-value} & \multicolumn{2}{c}{$<0.001$} \\
        \bottomrule
    \end{tabular}
\end{table}

\subsection{Human-Rater Study Results}

\subsubsection{Ratings in Isolation}

To establish global platform rankings that account for task complexity, we fitted Linear Mixed-Effects Models (LMM) to individual app ratings from isolated presentation stages. The model estimated platform quality scores while statistically controlling for three sources of variation: (1) prompt difficulty (random intercepts for each of 96 coding tasks), (2) participant grading bias (random intercepts for each participant to account for harsh vs. lenient raters), and (3) presentation order effects (fixed effect for position A vs. B). The model specification was:

\begin{equation}
\text{Score}_{ijk} = \beta_0 + \beta_{\text{platform}} + \beta_{\text{position}} + u_{i} + v_{j} + \epsilon_{ijk}
\end{equation}

where $u_i \sim \mathcal{N}(0, \sigma_u^2)$ represents random participant effects, $v_j \sim \mathcal{N}(0, \sigma_v^2)$ represents random prompt effects, and $\epsilon_{ijk}$ is the residual error.

For \textbf{Clarity}, all three platforms performed similarly after adjusting for difficulty, with Replit achieving a score of 3.96 (95\% CI: [3.82, 4.09]), Bolt at 3.95 (95\% CI: [3.82, 4.09]), and Firebase at 3.92 (95\% CI: [3.78, 4.06]). Pairwise differences were not statistically significant (Firebase vs. Bolt: $\beta = -0.037$, $p = 0.50$; Replit vs. Bolt: $\beta = 0.004$, $p = 0.94$).

\begin{figure}[htbp]
  \centering
  \includegraphics[width=.7\linewidth]{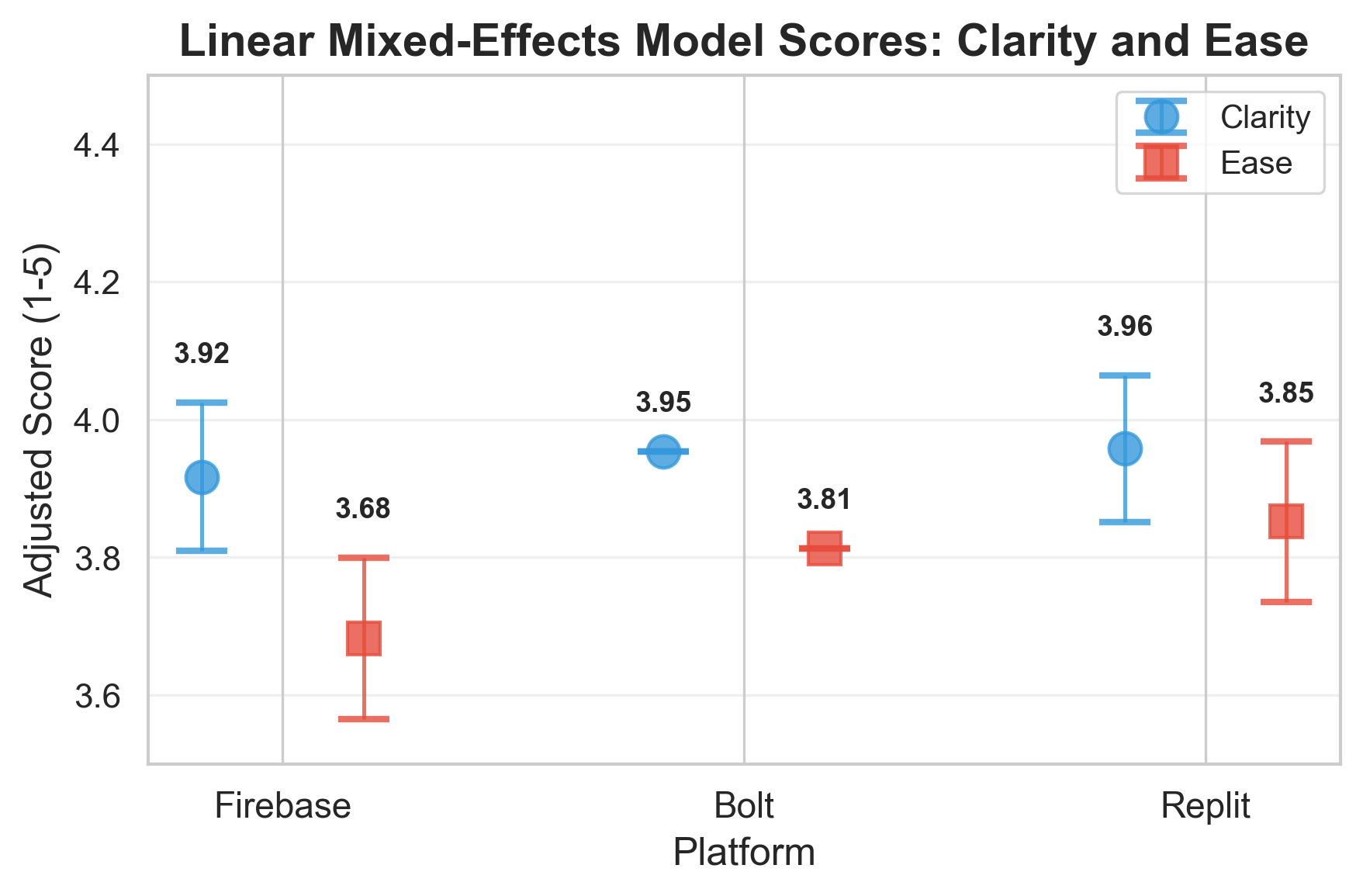}
  \caption{Combined Linear Mixed-Effects Model scores for Clarity and Ease metrics across all platforms. Scores represent ratings on a 1--5 scale, adjusted for participant bias, prompt difficulty, and position effects. Point plots with error bars show 95\% confidence intervals.}
  \label{fig:lmm_combined}
\end{figure}

\begin{figure}[htbp]
  \centering
  \includegraphics[width=\linewidth]{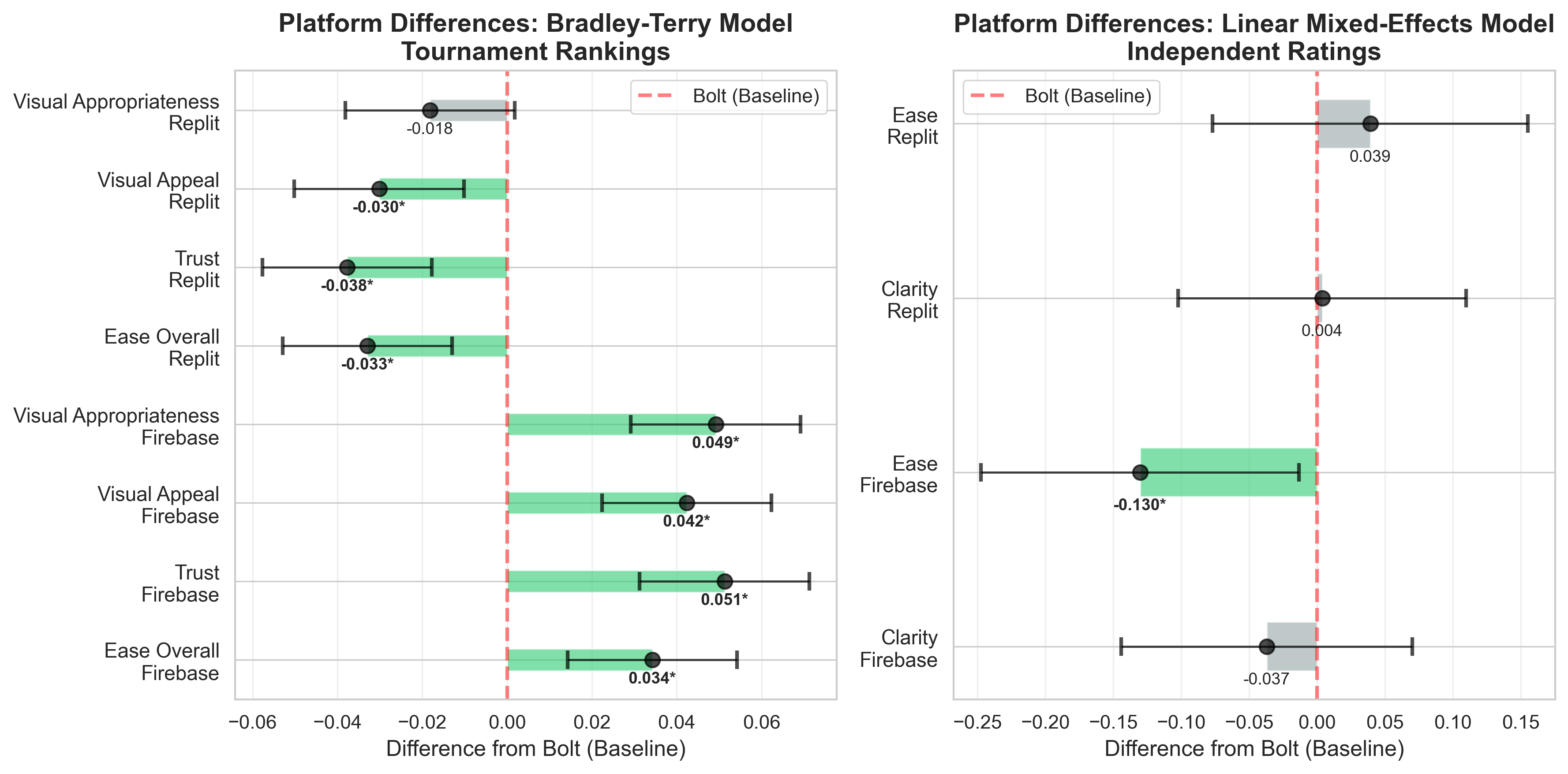}
  \caption{Platform performance differences from baseline (Bolt) measured using two independent statistical methodologies: Bradley-Terry tournament rankings (left) and Linear Mixed-Effects Models (right). Pairwise comparison results illustrate statistically significant differences in nearly all platform-dimension combinations. During isolated rating stages, a significant difference is only detected for one platform-dimension combination. This suggests that the pairwise comparison methodology is more sensitive to detecting user preferences than the isolated rating methodology. Error bars show 95\% confidence intervals. Asterisk (*) indicates statistical significance ($p < 0.05$).}
  \label{fig:platform_differences_combined}
\end{figure}

For \textbf{Ease of Use} (Figure \ref{fig:lmm_combined}), a slight but statistically significant difference in ranking emerged: Replit scored highest at 3.85 (95\% CI: [3.71, 3.99]), followed by Bolt at 3.81 (95\% CI: [3.67, 3.96]), and Firebase at 3.68 (95\% CI: [3.54, 3.82]). Firebase scored lower than Bolt ($\beta = -0.130$, $p = 0.029$), representing the only statistically meaningful difference in ease of use that we detected among the ratings in isolation. The Replit advantage over Bolt was positive but not statistically significant ($\beta = 0.039$, $p = 0.51$), see Figure \ref{fig:platform_differences_combined}.

\subsubsection{Pairwise Comparisons}

While isolated ratings provide a baseline for each system, the pairwise comparison stage reveals how systems compete when directly contrasted by the same user on the same task. This analysis is restricted to the 1,071 comparisons where both applications successfully appeared.

\paragraph{Preference Distributions and Win Rates}
We analyzed the distribution of user preferences across four key dimensions: Ease Overall, Trust, Visual Appeal, and Visual Appropriateness. As summarized in Table \ref{tab:combined_win_rates_bt}, Firebase achieved the highest win rates across all metrics, significantly outperforming competitors in all categories.

\begin{table*}[htbp]
\centering
\caption{Win rates and Bradley-Terry power rankings across all comparison metrics. Win rates represent the proportion of comparisons won by each platform with 95\% Wilson score confidence intervals. Bradley-Terry power scores represent log-odds of winning comparisons, with win probability vs average indicating the likelihood of beating an average opponent. Rankings are shown for each metric.}
\label{tab:combined_win_rates_bt}
\begin{tabular}{lcccccc}
\toprule
Platform & Metric & Win Rate & 95\% CI & Power Score & Win Prob. vs Avg & Rank \\
\midrule
Firebase & Ease Overall & 42.5\% & [0.388, 0.463] & 0.135 & 53.4\% & 1 \\
 & Trust & 41.2\% & [0.375, 0.450] & 0.187 & 54.7\% & 1 \\
 & Visual Appeal & 45.5\% & [0.418, 0.493] & 0.153 & 53.8\% & 1 \\
 & Visual Appropriateness & 45.1\% & [0.413, 0.489] & 0.156 & 53.9\% & 1 \\
\midrule
Bolt & Ease Overall & 31.7\% & [0.284, 0.352] & -0.002 & 50.0\% & 2 \\
 & Trust & 29.2\% & [0.259, 0.326] & -0.018 & 49.5\% & 2 \\
 & Visual Appeal & 36.8\% & [0.333, 0.404] & -0.016 & 49.6\% & 2 \\
 & Visual Appropriateness & 33.9\% & [0.306, 0.375] & -0.041 & 49.0\% & 2 \\
\midrule
Replit & Ease Overall & 26.0\% & [0.230, 0.292] & -0.134 & 46.7\% & 3 \\
 & Trust & 22.7\% & [0.199, 0.258] & -0.169 & 45.8\% & 3 \\
 & Visual Appeal & 28.9\% & [0.258, 0.322] & -0.137 & 46.6\% & 3 \\
 & Visual Appropriateness & 28.9\% & [0.258, 0.322] & -0.114 & 47.1\% & 3 \\
\bottomrule
\end{tabular}
\end{table*}

\paragraph{Tournament Rankings and Statistical Significance}
To synthesize these head-to-head results, we fitted Bradley-Terry models to estimate relative platform abilities (Table \ref{tab:combined_win_rates_bt}). This model treats each comparison as a tournament match, calculating the log-odds of a platform defeating an average opponent.

Firebase showed statistically significant advantages over Bolt in Trust ($\beta = 0.051$, $p < 0.05$), Visual Appropriateness ($\beta = 0.049$, $p < 0.05$), Visual Appeal ($\beta = 0.042$, $p < 0.05$), and Ease Overall ($\beta = 0.034$, $p < 0.05$). In contrast, Replit performed significantly worse than Bolt across most dimensions, particularly in Trust ($\beta = -0.038$) and Ease Overall ($\beta = -0.033$), see Figure \ref{fig:platform_differences_combined}.

\paragraph{Effect Sizes}
The magnitude of these differences was measured using Cliff's Delta ($\Delta$). As shown in the heatmap in Figure \ref{fig:effect_size_heatmap} (and Table \ref{tab:effect_sizes} in Appendix \ref{sec:appendix suplemental comparitive results}) the most pronounced differences favored Firebase over Bolt in Visual Appropriateness ($\Delta = -0.234$, small) and Visual Appeal ($\Delta = -0.212$, small). Comparisons between Firebase and Replit also revealed a small effect size in favor of Firebase for Trust ($\Delta = 0.198$).

\begin{figure}[htbp]
  \centering
  \includegraphics[width=0.8\linewidth]{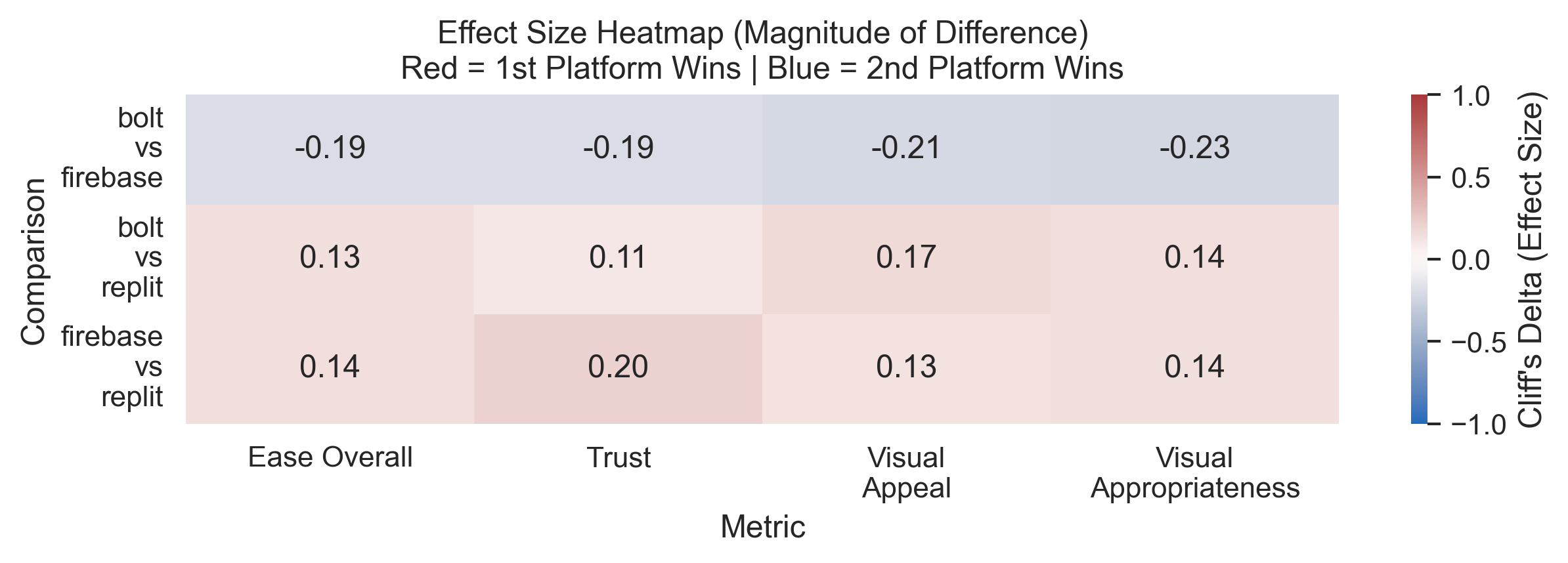}
  \caption{Effect sizes measured using Cliff\textquotesingle{}s Delta between platform pairs across comparison metrics. Effect size interpretation: $|\Delta| < 0.147$ (negligible), $0.147 \leq |\Delta| < 0.33$ (small), $0.33 \leq |\Delta| < 0.474$ (medium), $|\Delta| \geq 0.474$ (large). Color intensity indicates effect magnitude.}
  \label{fig:effect_size_heatmap}
\end{figure}

%% file: sections/5_discussion.tex
\section{Discussion and Limitations}
\label{sec:discussionlimitation}

The results of our benchmark provide an initial but detailed picture of the current landscape of prompt-to-app systems, revealing both progress and structural challenges. Several clear themes emerged across the automated metrics and human user evaluations, most notably systematic trade-offs between visual appeal, perceived usability, and user trust across platforms. 

\subsection{Discussion}

A prominent observation is that platform differences are more muted in isolated ratings, but become pronounced in head-to-head comparisons after task interaction. In isolation, \textbf{Clarity} scores are statistically indistinguishable across platforms (Table~\ref{tab:lmm_rankings}), and \textbf{Ease of Use} differences are small, with Firebase scoring slightly lower than Bolt ($\beta=-0.130$, $p=0.029$). In contrast, once users directly compared systems on the same task, Firebase consistently outperformed both Bolt and Replit across \textbf{Ease Overall}, \textbf{Trust}, \textbf{Visual Appeal}, and \textbf{Visual Appropriateness} (Table~\ref{tab:combined_win_rates_bt}; Fig.~\ref{fig:platform_differences_combined}). This stage-dependent separation may reflect a broader pattern in LLM-based software generation: superficial assessments can obscure comparative weaknesses that only emerge under more realistic interaction and task demands \cite{jimenez2024swebench}.

\paragraph{Participant Demographics.}

The human-rater study drew from a large and demographically heterogeneous participant pool, see Appendix~\ref{app:demographics} for more detail. This demographic breadth provides important context for interpreting the benchmark results. Although this study does not perform stratified or causal demographic analyses, the presence of consistent task-based patterns across a heterogeneous rater pool suggests that the reported differences in usability, trust, and perceived completeness are not driven solely by a single user subgroup. This interpretation is consistent with prior work showing that while user background can shape expectations and aesthetic preferences, task success and perceived reliability remain central drivers of user trust across populations \cite{li2024developing,paik2025who,cao2024designing}.

\subsection{Limitations}

While this benchmark represents an early and comprehensive attempt to evaluate prompt-to-app systems in a controlled, human-centered setting, several limitations shape the interpretation of our results and motivate future work.

\paragraph{Scope of Systems and Prompts.}
We evaluated three prompt-to-app systems across a curated set of 96 prompts spanning multiple domains, difficulty levels, specificity, and complexity. 
Although this prompt set is intentionally diverse, it does not cover the full space of real-world application requirements. 
As prompt-to-app systems mature, expanding the taxonomy of tasks and incorporating more domain-specific constraints will be essential for broader generalization.

\paragraph{Variability in Generative Outputs.}
Because prompt-to-app systems can produce different outputs across runs even for identical prompts, our evaluation reflects a single representative generation per system per prompt. 
This design choice mirrors existing LLM code generation benchmarks that rely on snapshot evaluations \cite{jimenez2024swebench,liu2023repobench}, but it limits our ability to characterize stochastic variability. 
Future work could incorporate multi-sampling, output stability metrics, or stochasticity-aware evaluation protocols to better capture generation consistency.

\paragraph{Automated Evaluation Coverage.}
The automated evaluation pipeline focused on deployability, framework usage, and basic correctness signals. 
It did not include full-stack dynamic testing, fuzzing, or runtime instrumentation. 
Prior research shows that surface-level checks often fail to detect deeper structural issues in AI-generated code, including latent race conditions, state inconsistencies, and unsafe defaults that emerge only during execution \cite{pearce2022asleep,yan2025guiding}. 
Accordingly, the automated metrics reported here should be interpreted as lower-bound indicators of functional robustness. 
Integrating dynamic analysis, type-flow consistency checks, and backend–frontend contract validation represents an important direction for future extensions of the benchmark.

\paragraph{Human-Rater Study Design.}
The human evaluation captures user perceptions following short, task-focused interactions with generated applications, but does not measure longer-term usability, learnability, accessibility, or maintainability. 
While participants represented a broad range of backgrounds and experience levels, the study was not designed to generalize to all user populations or usage contexts. 
Moreover, raters evaluated applications in controlled task blocks rather than through extended real-world use. 
Longitudinal studies, developer-oriented evaluations, and expert usability inspections would provide complementary perspectives beyond the scope of the present work.

\paragraph{Interpretation of Aesthetic Preferences.}
Although aesthetic preferences were measured consistently across raters, design judgments may vary in real-world contexts depending on cultural norms, industry conventions, or brand expectations. 
Our evaluation focuses on general-purpose perceptions of visual appeal and appropriateness rather than domain-specific design criteria. 
Future work could explore domain-tailored aesthetic evaluations or incorporate expert design review to assess alignment with professional UI standards.

%% file: sections/6_conclusion.tex
\section{Conclusion and Future Work}
\label{sec:conclusion}

We have presented a large-scale, human-centered benchmark for evaluating prompt-to-app agentic systems. By combining automated analysis with a rigorous, task-based human evaluation, we assessed three widely used commercial systems and assessed their strengths and limitations from a user-centered perspective.

The comparative evaluation of agentic ``prompt-to-app'' systems reveals a landscape of high absolute performance with distinct relative preferences. While the systems are not strictly interchangeable, they all demonstrate high utility, with absolute ``Clarity'' and ``Ease of Use'' scores between 3.68 and 3.96 on a 5-point scale.

Statistical analysis shows that platform performance is highly dependent on the comparison stage. In isolated assessments, differences in Clarity were statistically indistinguishable (p>.5), and Replit achieved the highest absolute score for Ease of Use (3.85 out of 5). In pairwise comparisons though, where both applications successfully deployed, Firebase Studio emerged as the top-ranked system across all metrics:

\begin{itemize}
\item \textbf{Interactive Ease:} Firebase secured a 42.5\% win rate, significantly outperforming Bolt (31.7\%) and Replit (26.0\%).

\item \textbf{User Trust:} Firebase won 41.2\% of head-to-head matchups, compared to 29.2\% for Bolt and 22.7\% for Replit.

\item \textbf{Visual Dimensions:} Firebase maintained a lead in Visual Appeal (45.5\%) and Visual Appropriateness (45.1\%), with Bolt remaining competitive at (36.8\%) and (33.9\%) respectively.
\end{itemize}

These results establish a clear performance hierarchy (Firebase $>$ Bolt $>$ Replit) when users move beyond static observation to functional interaction. Firebase Studio achieved the highest composite score (1.933), driven by superior win rates in trust and interactive ease, followed by Bolt (1.561) then Replit (1.302).

Our results yield two primary conclusions regarding the current state of agent ``prompt-to-app'' systems. First, \textbf{independent assessments are statistically indistinguishable}, suggesting that raters struggle to provide reliable reference-free evaluations in this domain. Consequently, \textbf{pairwise (side-by-side) comparisons are the only reliable methodology for establishing a meaningful performance benchmark}. Second, the current ecosystem does not require a trade-off between key attributes; \textbf{a single system (Firebase Studio) significantly outperformed competitors across all measured axes}, including trust, ease of use, and visual appeal. This indicates that the market is not yet commoditized and certain systems currently dominate across the board, though this may shift as the technology evolves.

Looking ahead, our results suggest several concrete directions for future work:

\begin{itemize}
    \item \textbf{Expanded Benchmark Coverage.}
    We plan to extend the prompt set to include more complex workflows, multi-user roles, and integrations with third-party APIs, as well as prompts that stress long-lived state, permissions, and multi-step transactions.

    \item \textbf{Stability and Multi-Sample Evaluation.}
    Future iterations will explore multi-sampling for each system--prompt pair, enabling direct measurement of output variance, robustness to prompt perturbations, and consistency across generations.

    \item \textbf{Stronger Security and Reliability Analysis.}
    We aim to incorporate richer static and dynamic evaluation pipelines, including runtime testing and security-oriented analyses, to better characterize the reliability and risk profile of generated applications, building on prior work highlighting limitations of surface-level checks \cite{pearce2022asleep}.

    \item \textbf{Developer-Centered and Longitudinal Studies.}
    Beyond end-user ratings, we plan to study how developers interact with generated applications over longer time horizons—modifying, extending, and debugging them—to quantify \emph{agentic ROI} in terms of time-to-prototype, maintenance effort, and reduction in developer workload.
\end{itemize}

Finally, we intend to release the benchmark framework, generated artifacts, and survey instruments to support reproducible evaluation and enable the community to extend the benchmark with new systems, tasks, and metrics over time.

The prompt-to-app paradigm is still in its early stages. We believe that rigorous, human-centered benchmarks such as this one are essential for guiding the development of agentic systems toward tools that are not only visually compelling, but also functionally reliable, trustworthy, and practically useful.

%% file: sections/7_appendix.tex
\subsection{Artifact Generation Details}
\label{subsec:artifact_generation_details}

\subsubsection{ Firebase Studio}
Artifact generation for Firebase Studio was fully automated using a custom browser automation script built with Puppeteer, a Node.js library for controlling headless Chrome. Apps were generated on or around September 5, 2025 using a paid Firebase Studio account.

The automation script executed a multi-phase workflow for each prompt:
Initialization: The script launched a Chrome browser instance with persistent session storage to maintain authentication state. Prior to processing, the script verified GitHub API permissions by issuing authenticated requests to ensure the target repository owner (user or organization) had valid credentials for repository creation.
Prompt Submission: For each prompt, the script navigated to the Firebase Studio workspace interface, located the prompt input field using CSS selectors, and programmatically entered the prompt text. Prompts were submitted by simulating a keyboard ``Enter'' event. The script then waited for the platform to generate an initial application blueprint.
Prototype Generation: After blueprint generation (typically 25 seconds), the script searched for and clicked the ``Prototype this App'' button within the Firebase Studio interface. This triggered the platform's full application generation process. The script monitored for completion indicators for up to 60 attempts at 2-second intervals.
Code Export: Once generation completed, the script transitioned the workspace to developer mode, opened an integrated terminal via the VS Code command palette (F1 → ``Terminal: Create New Terminal''), and executed a sequence of Git commands to initialize a repository, configure credentials, and push the generated code to a pre-created GitHub repository.
Workspace Cleanup: Upon successful export verification (confirmed via GitHub API commit check), the script deleted the Firebase Studio workspace to avoid accumulating workspace clutter and to maintain consistent state for subsequent generations.
Concurrency and Rate Limiting. The script processed prompts in batches of 2 concurrent tabs to balance throughput against platform rate limits. A staggered delay was introduced between prompt submissions within each batch. When rate limit responses were detected (identified by parsing page content for ``rate limit'' or ``too many requests'' messages), the script paused for a 3-minute cooldown period before retrying.
Error Handling and Recovery. The automation included retry logic for transient failures. If the ``Prototype this App'' button was not found within the expected timeframe—often indicating a ``bugged'' workspace state—the script attempted workspace recovery by navigating back to the main workspace page, deleting the problematic workspace, and retrying generation with a fresh workspace. Each prompt was allowed up to 2 workspace recovery attempts before being marked as a persistent failure requiring manual investigation.
Execution Details. Apps were generated on or around September 5, 2025 using a paid Firebase Studio account. The automation ran in non-headless mode to allow for initial manual authentication. Generated artifacts were pushed to GitHub repositories under a dedicated organization, with repository names derived from sanitized prompt text. A JSON results file was generated at the end of each run, recording timestamps, success/failure status, workspace IDs, and any errors encountered for each prompt.

\subsubsection{ Bolt}
Artifacts generation was triggered manually via the Bolt web interface, on or around September 28, 2025. Artifacts were pushed to GitHub manually using the Bolt web interface.

In a small number of instances, during app generation, the app building agent would ask for additional user input in the form of a binary choice. In these instances one of the choices was selected at random and the app builder was allowed to proceed until the app was completed.

\subsubsection{ Replit}
Artifacts generation was triggered manually via the Replit web interface, on or around September 28, 2025. Artifacts were pushed to GitHub manually using the Replit web interface.

In a small number of instances, during app generation, the app building agent would ask for additional user input in the form of an API key, along with instructions for obtaining the requested key. In these instances the user followed the instructions as provided and entered the key into the interface, and the app builder was allowed to proceed until the app was completed.

\subsection{Artifact Deployment Details}
\label{subsec:artifact_deployment_details}
Deployment of generated artifacts was performed largely automatically, with occasional need for manual intervention. In particular, some Replit artifacts required a manual step to add API keys to Quome Cloud for injection into the app. Additionally, in cases where apps expected a database connection, a volume hosting a SQLite database was manually added to the app specifications in Quome Cloud.

A helper program was developed to automatically deploy generated artifacts through a four-step process: checking out the artifacts from GitHub, detecting the application type and stack, building a Docker image, and finally creating an app specification for deployment in Quome Cloud.

\paragraph{Artifact Checkout and Triage.}
Artifacts were pulled from GitHub using the GitHub REST API. Depending on the artifact's source (Firebase, Replit, Bolt, Quome), an additional step was performed to determine the app stack. Artifacts generated by Firebase, Quome, and Bolt possessed a consistent structure and required no additional triage. However, artifacts from Replit showed variation. A few Bolt and Replit artifacts contained no code due to generation errors; in these cases, subsequent steps were skipped. All remaining Replit artifacts were sorted into four categories: static web pages (HTML and JavaScript only), apps served by a Python backend, apps served by a Node.js backend, and apps served with Vite.

\paragraph{Deployment to Quome Cloud.}
After sorting the application types, a corresponding Dockerfile was added to each artifact repository to ensure the correct startup commands were run. Quome-generated artifacts came with a Dockerfile, which was not replaced during this step. Additionally, a simple GitHub Actions workflow was added, and the secrets required to access the container registry and Quome Cloud API were configured in the GitHub repository. This workflow builds a Docker image from the artifact, pushes it to a container registry, and pushes the updated application specification to Quome Cloud.

\subsection{Automated Audit of Deployments}
\label{sec:appedix automated audit}

To validate that participant ratings reflect actual deployment outcomes rather than random responses, we conducted a comprehensive comparison between automated deploy testing and participant-reported appearance rates.

Table~\ref{tab:deploy_agreement_summary} presents the overall agreement statistics. We found substantial agreement between automated testing and participant reports, with an overall agreement rate of 88.2\% (254 of 288 URLs). Cohen's Kappa, which accounts for chance agreement, was 0.691, indicating substantial inter-rater reliability. The Pearson correlation coefficient was 0.706 ($p < 0.001$), demonstrating a strong positive linear relationship between automated success rates and participant appeared rates.

\begin{table}[htbp]
    \centering
    \caption{Agreement between automated deploy testing and participant-reported appearance rates. Cohen's Kappa accounts for chance agreement. Correlation measures linear relationship between automated success and participant appeared rates.}
    \label{tab:deploy_agreement_summary}
    \begin{tabular}{lc}
        \toprule
        \textbf{Metric} & \textbf{Value} \\
        \midrule
        URLs Compared & 288 \\
        Agreement Rate & 88.2\% \\
        Mean Absolute Difference & 0.252 \\
        Pearson Correlation & 0.706 \\
        Correlation $p$-value & $<0.001$ \\
        Cohen's Kappa & 0.691 \\
        Kappa Interpretation & Substantial \\
        \bottomrule
    \end{tabular}
\end{table}

Table~\ref{tab:deploy_platform_comparison} presents platform-specific comparisons. Agreement rates varied somewhat across platforms: Replit showed the highest agreement (92.7\%), followed by Bolt (90.6\%), and Firebase (81.2\%). The mean absolute difference between automated success rates and participant appeared rates was smallest for Replit (0.232) and largest for Firebase (0.282).

Notably, Firebase showed the largest discrepancy: automated testing reported 84.4\% success rate, while participants reported only 68.2\% appeared rate—a difference of 16.2 percentage points. This suggests that Firebase deployments may have issues detectable by participants but not captured by automated testing (e.g., slow loading, rendering problems, or usability issues). In contrast, Bolt showed nearly identical rates (64.6\% automated vs. 64.8\% participant), indicating excellent agreement for this platform.

Table~\ref{tab:deploy_agreement_by_platform} provides detailed counts of agreements and disagreements by platform. Overall, 254 URLs showed agreement (88.2\%) while 34 showed disagreement (11.8\%).

\begin{table}[htbp]
    \centering
    \caption{Platform-specific comparison of automated deploy success rates vs. participant-reported appearance rates. Agreement rate shows percentage of URLs where automated and participant results match. Mean absolute difference quantifies the average discrepancy.}
    \label{tab:deploy_platform_comparison}
    \begin{tabular}{lccccc}
        \toprule
        \textbf{Platform} & \textbf{URLs} & \textbf{Automated} & \textbf{Participant} & \textbf{Agreement} & \textbf{Mean Abs. Diff.} \\
        & & \textbf{Success} & \textbf{Appeared} & \textbf{Rate} & \\
        \midrule
        Bolt & 96 & 64.6\% & 64.8\% & 90.6\% & 0.244 \\
        Firebase & 96 & 84.4\% & 68.2\% & 81.2\% & 0.282 \\
        Replit & 96 & 75.0\% & 68.4\% & 92.7\% & 0.232 \\
        \bottomrule
    \end{tabular}
\end{table}

\begin{table}[htbp]
    \centering
    \caption{Detailed agreement statistics by platform showing number of URLs with agreement and disagreement between automated testing and participant reports.}
    \label{tab:deploy_agreement_by_platform}
    \begin{tabular}{lcccc}
        \toprule
        \textbf{Platform} & \textbf{Total URLs} & \textbf{Agreements} & \textbf{Disagreements} & \textbf{Agreement Rate} \\
        \midrule
        Bolt & 96 & 87 & 9 & 90.6\% \\
        Firebase & 96 & 78 & 18 & 81.2\% \\
        Replit & 96 & 89 & 7 & 92.7\% \\
        \midrule
        \textbf{Overall} & \textbf{288} & \textbf{254} & \textbf{34} & \textbf{88.2\%} \\
        \bottomrule
    \end{tabular}
\end{table}

\subsection{ Cross-System Standardization}
To ensure fair comparison across systems with different default behaviors, we applied the following standardization procedures:

\paragraph{Prompt Delivery.}
All prompts were submitted verbatim without modification. No system-specific prompt engineering or optimization was performed. Prompts were delivered through each system's default web interface using the primary text input field.
Configuration Defaults. All systems were used with their default configuration settings. No optional features (such as specifying a technology stack, enabling/disabling specific frameworks, or adjusting generation parameters) were modified from their out-of-box state.

\paragraph{Single-Shot Generation.} Each artifact was generated in a single pass from the initial prompt. We did not use iterative refinement, follow-up prompts, or manual code editing. If a system's agent requested clarifying input during generation (as occurred with Bolt and Replit), one of the offered choices was selected at random to simulate a minimally-engaged user.

\paragraph{Temporal Consistency.} To minimize the impact of model updates or platform changes, all artifacts for each system were generated within a concentrated time window (see system-specific subsections). We acknowledge that prompt-to-app systems are rapidly evolving, and results may differ with subsequent model versions.

\subsection{Filtering for Quality Responses}
\label{appendix:filtering}

We wanted to ensure that the results we included in our analysis were more likely to come from participants who were deliberate with their responses, rather than making sections at random with the sole intention of getting credit for completing the survey. To do this, we implemented a simple smoke test, and a set of quality gates based on observations of participants behavior tracked by our survey page analytics.

\paragraph{Basic Quality Filters}

We imposed two initial filters. First, we only included participants who spent at least 12.5 minutes on completing the survey, which we felt was an indicator that they had issues that resulted in them giving up or not completing the survey.  Then we discarded participants who completed 20 or more comparisons, which we felt indicated a survey taker responding too quickly to meaningfully consider their responses.

\paragraph{Smoke Test}

All survey participants were presented with the same first task:
\begin{enumerate}
    \item \textbf{Page 1: Isolated Task for App A.}  
    The rater is shown the front page of Wikipedia and given a specific functional task: ``Search for information about Albert Einstein.''  

    \item \textbf{Page 2: Isolated Task for App B.}  
    The rater is shown a page that fails to load and is asked to perform the exact same task.  

    \item \textbf{Page 3: Side by Side Preference.}  
    The rater is shown these pages side by side and asked to complete their ratings.
\end{enumerate}

We discarded all responses from raters who did not provide the expected responses for these tasks:

\begin{enumerate}
    \item \textbf{Page 1: Isolated Task for App A, expected responses}: ``Yes'' the app appeared, the clarity of the apps purpose should be neutral to positive, and the ease of the task should be neural to positive.

    \item \textbf{Page 2: Isolated Task for App B, expected responses: } ``No'' the app did not appear. We did not consider the ratings for the clarity of the app's purpose or the ease of the task in the smoke test.  

    \item \textbf{Page 3: Side by Side Preference, expected responses:} Report that ``Only A'' appeared. We did not consider comparative ratings in the smoke test.
\end{enumerate}

Responses that did not meet these criteria indicated to us that the participant either misunderstood the response format or otherwise did not follow the instructions. In combination with the basic quality filters, this smoke test left us with 205 participants, having completed 1,986 comparisons, see Table \ref{tab:filter_cascade}.

\begin{table}[htbp]
\centering
\caption{Participant Filter Cascade: Retention at Each Stage, and Average Number of Comparisons per Participant}
\label{tab:filter_cascade}
\begin{tabular}{lrrrr}
\toprule
\textbf{Filter Stage} & \textbf{Participants} & \textbf{Retention (\%)} & \textbf{Comparisons} & \textbf{Avg/Participant} \\
\midrule
All Participants & 828 & 100.0 & 4,410 & 5.3 \\
Base Filters & 404 & 48.8 & 3,751 & 9.3 \\
+ Smoke Test & 205 & 24.8 & 1,986 & 9.7 \\
\bottomrule
\end{tabular}
\end{table}

\begin{table}[htbp]
\centering
\caption{Isolated Stage Appearances by Platform}
\label{tab:isolated_by_platform}
\small
\begin{tabular}{lrrr}
\toprule
\textbf{Platform} & \textbf{All Participants} & \textbf{Base Filters} & \textbf{+ Smoke Test} \\
\midrule
Bolt & 2,430 (73.7\%) & 2,065 (73.4\%) & 982 (66.6\%) \\
Firebase & 2,289 (75.5\%) & 1,959 (75.2\%) & 918 (69.1\%) \\
Replit & 2,627 (75.6\%) & 2,246 (75.2\%) & 1,053 (70.2\%) \\
\midrule
\textbf{Total} & \textbf{7,346 (74.9\%)} & \textbf{6,270 (74.6\%)} & \textbf{2,953 (68.6\%)} \\
\textbf{Out of} & \textit{9,803} & \textit{8,407} & \textit{4,304} \\
\bottomrule
\end{tabular}
\vspace{0.5em}
\begin{minipage}{\linewidth}
\small
\textit{Note:} Combines both Isolated A and Isolated B stages. Values show number of successful appearances with percentage in parentheses. ``Out of'' row shows total isolated stage presentations.
\end{minipage}
\end{table}

\begin{table}[htbp]
\centering
\caption{Platform Pair Comparison Statistics: All Comparisons and Both-Appeared Only}
\label{tab:platform_pairs}
\small
\begin{tabular}{lrrrrrr}
\toprule
& \multicolumn{3}{c}{\textbf{All Comparisons}} & \multicolumn{3}{c}{\textbf{Both Appeared}} \\
\cmidrule(lr){2-4} \cmidrule(lr){5-7}
\textbf{Platform Pair} & \textbf{All} & \textbf{Base} & \textbf{+Smoke} & \textbf{All} & \textbf{Base} & \textbf{+Smoke} \\
\midrule
Bolt vs Firebase & 1,379 & 1,157 & 615 & 825 & 694 & 309 \\
Bolt vs Replit & 1,603 & 1,370 & 740 & 997 & 848 & 401 \\
Firebase vs Replit & 1,428 & 1,224 & 631 & 881 & 756 & 361 \\
\midrule
\textbf{Total} & \textbf{4,410} & \textbf{3,751} & \textbf{1,986} & \textbf{2,703} & \textbf{2,298} & \textbf{1,071} \\
\textbf{Both-appeared rate} & \multicolumn{3}{c}{---} & \textbf{61.3\%} & \textbf{61.3\%} & \textbf{53.9\%} \\
\bottomrule
\end{tabular}
\vspace{0.5em}
\begin{minipage}{\linewidth}
\small
\textit{Note:} ``All Comparisons'' includes all comparison stage records regardless of appearance. ``Both Appeared'' includes only comparisons where both apps successfully appeared.
\end{minipage}
\end{table}

\subsection{Survey Details}
\label{appendix:survey_details}

\paragraph{Participant Demographics}

\begin{figure}[htbp]
  \centering
  \includegraphics[width=\linewidth]{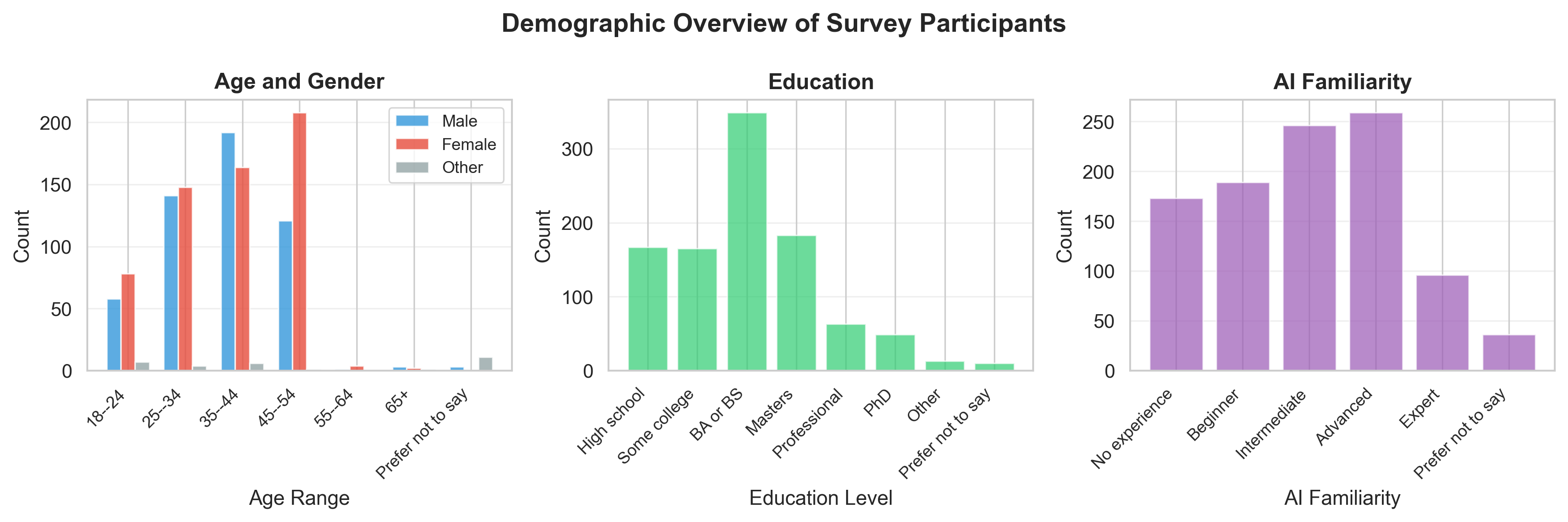}
  \caption{Demographic characteristics of survey participants showing age distribution by gender (Male, Female, Other), education levels, and AI familiarity. This chart represents all participants who provided demographic data. Filtered participant demographics are not available due to anonymization constraints.}
  \label{fig:demographic_overview}
\end{figure}

Figure~\ref{fig:demographic_overview} summarizes the demographic characteristics of participants who contributed human evaluations. 
The rater pool spans a broad range of ages, education levels, professional backgrounds, and familiarity with AI and software development tools. 
Participants include both technical and non-technical users, with substantial representation from software and technology, healthcare, education, and other domains.

We report these distributions to contextualize the human evaluation results. 
We do not perform subgroup analyses by demographic category in this work.

\begin{table*}[htbp]
\centering
\caption{Demographic characteristics of survey participants (N = 999). This table represents all participants who provided demographic data. Filtered participant demographics are not available due to anonymization constraints that prevent matching demographic records to filtered participant IDs. All demographic fields had complete data with no missing values.}
\label{tab:demographics}
\begin{tabular}{llr}
\toprule
\textbf{Category} & \textbf{Value} & \textbf{Count (\%)} \\
\midrule
 \multirow{7}{*}{\textbf{Age Range}} & 35--44 & 324 (32.4\%) \\
  & 45--54 & 301 (30.1\%) \\
  & 25--34 & 233 (23.3\%) \\
  & 18--24 & 123 (12.3\%) \\
  & 55--64 & 3 (0.3\%) \\
  & Prefer not to say & 10 (1.0\%) \\
  & 65+ & 5 (0.5\%) \\
\midrule
 \multirow{5}{*}{\textbf{Gender}} & Woman & 532 (53.3\%) \\
  & Man & 448 (44.8\%) \\
  & Non-binary & 9 (0.9\%) \\
  & Prefer not to say & 9 (0.9\%) \\
  & Prefer to self-describe & 1 (0.1\%) \\
\midrule
 \multirow{8}{*}{\textbf{Education Level}} & Bachelor's degree & 349 (34.9\%) \\
  & Master's degree & 183 (18.3\%) \\
  & High school or equivalent & 167 (16.7\%) \\
  & Some college / undergraduate studies & 165 (16.5\%) \\
  & Professional degree (MD, JD, MBA, etc.) & 63 (6.3\%) \\
  & Doctoral degree & 49 (4.9\%) \\
  & Other & 13 (1.3\%) \\
  & Prefer not to say & 10 (1.0\%) \\
\midrule
 \multirow{10}{*}{\textbf{Top Industries}} & Software / Technology & 316 (31.6\%) \\
  & Other & 183 (18.3\%) \\
  & Prefer not to say & 106 (10.6\%) \\
  & Healthcare / Health Tech & 95 (9.5\%) \\
  & Finance / FinTech & 86 (8.6\%) \\
  & Education & 75 (7.5\%) \\
  & Operations / HR / Management & 55 (5.5\%) \\
  & Arts / Media / Creative fields & 36 (3.6\%) \\
  & Government / Public Sector & 26 (2.6\%) \\
  & Research / Academia & 21 (2.1\%) \\
\midrule
 \multirow{6}{*}{\textbf{Years of Experience}} & 15+ years & 177 (17.7\%) \\
  & Prefer not to say & 81 (8.1\%) \\
  & 0--1 years & 82 (8.2\%) \\
  & 2--4 years & 183 (18.3\%) \\
  & 5--9 years & 267 (26.7\%) \\
  & 10--14 years & 209 (20.9\%) \\
\midrule
 \multirow{6}{*}{\textbf{AI Familiarity}} & No experience & 173 (17.3\%) \\
  & Beginner & 189 (18.9\%) \\
  & Intermediate & 246 (24.6\%) \\
  & Advanced & 259 (25.9\%) \\
  & Expert / Professional developer & 96 (9.6\%) \\
  & Prefer not to say & 36 (3.6\%) \\
\bottomrule
\end{tabular}
\end{table*}

\paragraph{Refined Task Based Design.}
We developed a structured, multi stage evaluation flow. The goal of the revised design was to ensure that raters engaged more deeply with each application's functionality in isolation before making comparative judgments. 

Each participant evaluated multiple pairs of applications through a three-phase process. First, participants interacted with App A in isolation, rating its clarity of purpose (1=very unclear, 5=very clear) and ease of task completion (1=very difficult, 5=very easy). Second, participants independently evaluated App B using identical measures. Finally, participants viewed both applications side-by-side and provided comparative judgments across four dimensions: visual appeal, visual appropriateness for the task, overall ease of use, and trustworthiness. All comparison questions used a 5-point preference scale (1=strongly prefer A, 3=no preference, 5=strongly prefer B), see Table \ref{tab:survey_design}. This quantitative study design, as opposed to open-ended qualitative feedback, adopts the human-centered evaluation protocols used in clinical evaluation~\citep{patient-centric}

\begin{table}[htbp]
\centering
\caption{Survey Design: Question Structure and Measurement Scales}
\label{tab:survey_design}
\small
\begin{tabularx}{\linewidth}{p{2.2cm} X p{3.5cm}}
\toprule
\textbf{Stage \& Category} & \textbf{Question \& Scale} & \textbf{Purpose} \\
\midrule
\multicolumn{3}{l}{\textbf{Isolated Stage (Individual Evaluation)}} \\
\midrule
Appearance & \textit{Did the app appear on screen?} \newline Yes / No & Technical validation \\
\addlinespace
Clarity & \textit{How clear is the app's purpose?} \newline 1--5 (very unclear to very clear) & Measure perceived clarity of purpose \\
\addlinespace
Ease of Use & \textit{How easy or difficult was this task?} \newline 1--5 (very difficult to very easy) & Measure task completion difficulty \\
\midrule
\multicolumn{3}{l}{\textbf{Comparison Stage (Side-by-Side Evaluation)}} \\
\midrule
Appearance & \textit{Did both apps appear on screen?} \newline Both / Only A / Only B & Technical validation \\
\addlinespace
Visual Appeal & \textit{Which app has a more appealing visual style?} \newline 1--5 preference scale & Measure aesthetic preferences \\
\addlinespace
Visual Appropriateness & \textit{Which app has a more appropriate visual style for the task?} \newline 1--5 preference scale & Measure design appropriateness \\
\addlinespace
Comparative Ease & \textit{Now that you have used both, which app was easier to use overall?} \newline 1--5 preference scale & Measure relative ease of use \\
\addlinespace
Trust & \textit{Which app did you trust more?} \newline 1--5 preference scale & Measure perceived trust \\
\bottomrule
\end{tabularx}
\vspace{0.3em}
\begin{minipage}{\linewidth}
\small
\textit{Note:} Comparison stage preference scale: 1=strongly prefer A, 3=no preference, 5=strongly prefer B. All questions required for complete comparison.
\end{minipage}
\end{table}

\begin{enumerate}
    \item \textbf{Page 1: Isolated Task for App A.}  
    The rater is shown only App A, embedded in an iframe, and given a specific functional task.  
    For example, in the medication app study:  
    \emph{Using the app below, please add “Lisinopril” for 9:00 AM.}  
    After completing the task, the rater evaluates clarity and task difficulty.

    \item \textbf{Page 2: Isolated Task for App B.}  
    The rater is shown only App B and performs the exact same task.  
    This produces a direct, task level comparison of functional completeness.

    \item \textbf{Page 3: Side by Side Preference.}  
    After using both apps, the rater is shown them side by side and asked holistic preference questions regarding visual appeal, ease of use, completeness, and trust.
\end{enumerate}

This design offers two key advantages. First, isolated evaluations capture absolute quality perceptions without comparison bias, providing baseline measures of clarity and usability. Second, direct comparisons reveal relative preferences on subjective dimensions like visual appeal and trust, where comparative judgment may be more reliable than absolute ratings. At both stages, we recorded whether applications successfully appeared on screen, allowing us to account for technical failures in our analysis.

Participants completed as many app pair comparisons as they chose within the survey time limit, 15 minutes. We required all questions within each stage to be answered for a comparison to be considered complete and included in the analysis.

\subsection{Statistical Methodology Details}
\label{appendix:stats}

For hypothesis testing, pairwise platform comparisons used the Wilcoxon signed-rank test. For isolated metrics comparing multiple platforms simultaneously, we applied the Kruskal-Wallis H test to assess overall differences across platforms, and when sufficient paired data existed, we used the Friedman test for repeated measures. To account for multiple comparisons, we applied the Benjamini-Hochberg False Discovery Rate (FDR) correction procedure across all hypothesis tests at $\alpha = 0.05$, reporting both uncorrected p-values and FDR-corrected significance.

Beyond hypothesis testing, we estimated effect sizes using Cliff's Delta. We calculated 95\% confidence intervals using the Wilson score method for proportions (win rates, appearance rates) and the t-distribution for continuous measures (mean ratings). To establish global platform rankings while controlling for confounds, we fitted two complementary models: (1) Linear Mixed-Effects Models (LMM) with platform and presentation position as fixed effects and random intercepts for participants (to control for harsh vs. lenient grader bias) and prompts (to control for task difficulty), and (2) Bradley-Terry models treating the evaluation as a pairwise tournament to estimate relative platform abilities. We also conducted post-hoc statistical power analyses to assess the adequacy of our sample sizes for detecting meaningful differences.

\subsection{Supplemental Isolated Results}

This appendix provides additional statistical charts and visualizations supporting the isolated response results discussed in Section \ref{sec:results}. These metrics further illustrate the performance disparities between Firebase, Bolt, and Replit across trust and aesthetic dimensions.

\begin{table}[htbp]
    \centering
    \caption{Absolute Quality Scores from Linear Mixed-Effects Models controlling for participant bias, prompt difficulty, and position effects. Scores represent ratings on a 1--5 scale. Baseline platform is Bolt. $\Delta$ represents difference from baseline. Asterisk (*) indicates statistical significance ($p < 0.05$).}
    \label{tab:lmm_rankings}
    \begin{tabular}{lccccc}
        \toprule
        Platform & Metric & Score & 95\% CI & $\Delta$ vs Baseline & $p$-value \\
        \midrule
        Replit & Clarity & 3.96 & [3.85, 4.06] & 0.004 & $0.944$ \\
        Bolt &  & 3.95 & [3.82, 4.09] & --- & $1.000$ \\
        Firebase &  & 3.92 & [3.81, 4.02] & -0.037 & $0.500$ \\
        \midrule
        Replit & Ease & 3.85 & [3.74, 3.97] & 0.039 & $0.508$ \\
        Bolt &  & 3.81 & [3.67, 3.96] & --- & $1.000$ \\
        Firebase &  & 3.68 & [3.57, 3.80] & -0.130 & $0.029$* \\
        \bottomrule
    \end{tabular}
\end{table}

\subsection{Supplemental Comparative Results}
\label{sec:appendix suplemental comparitive results}

This appendix provides additional statistical tables and visualizations supporting the pairwise comparison results discussed in Section \ref{sec:results}. These metrics further illustrate the performance disparities between Firebase, Bolt, and Replit across trust and aesthetic dimensions.

\subsubsection{Effect Size Across Comparison Metrics}

Table \ref{tab:effect_sizes} details the results shown in Figure \ref{fig:effect_size_heatmap}.

\begin{table}[htbp]
    \centering
    \caption{Effect sizes measured using Cliff's Delta between platform pairs across comparison metrics. Effect size interpretation: $|\Delta| < 0.147$ (negligible), $0.147 \leq |\Delta| < 0.33$ (small), $0.33 \leq |\Delta| < 0.474$ (medium), $|\Delta| \geq 0.474$ (large). Positive values indicate the first platform performs better, negative values indicate the second platform performs better.}
    \label{tab:effect_sizes}
    \begin{tabular}{lcc}
        \toprule
        Comparison & Metric & Cliff's $\Delta$ \\
        \midrule
        Bolt vs Firebase & Visual Appropriateness & -0.234 (small) \\
        Bolt vs Firebase & Visual Appeal & -0.212 (small) \\
        Firebase vs Replit & Trust & 0.198 (small) \\
        Bolt vs Firebase & Trust & -0.191 (small) \\
        Bolt vs Firebase & Ease Overall & -0.188 (small) \\
        Bolt vs Replit & Visual Appeal & 0.166 (small) \\
        Bolt vs Replit & Visual Appropriateness & 0.142 (negligible) \\
        Firebase vs Replit & Ease Overall & 0.141 (negligible) \\
        Firebase vs Replit & Visual Appropriateness & 0.135 (negligible) \\
        Bolt vs Replit & Ease Overall & 0.135 (negligible) \\
        Firebase vs Replit & Visual Appeal & 0.126 (negligible) \\
        Bolt vs Replit & Trust & 0.107 (negligible) \\
        \bottomrule
    \end{tabular}
\end{table}

\subsubsection{Distributions of Comparison Outcomes}

\begin{figure}[htbp]
  \centering
  \includegraphics[width=0.8\linewidth]{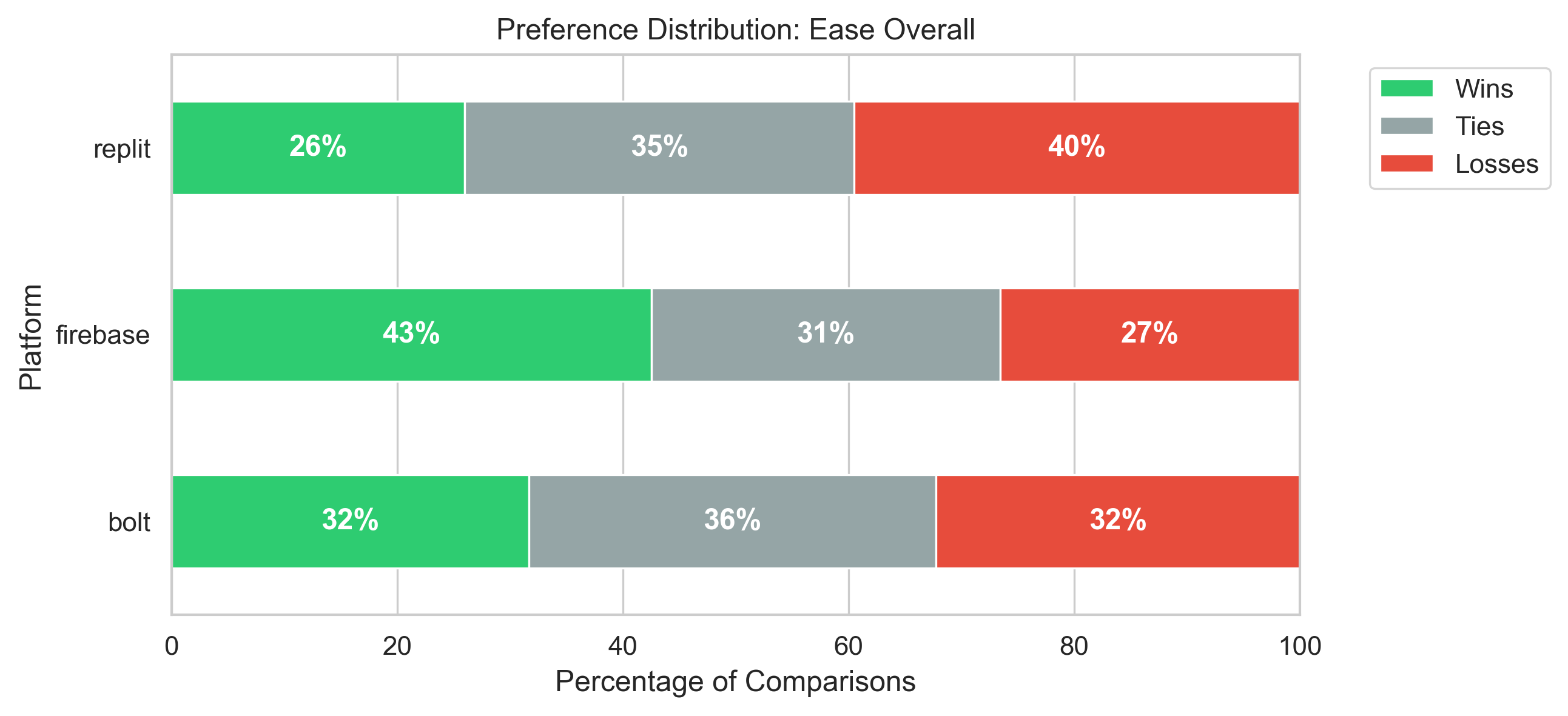}
  \caption{Distribution of comparison outcomes (Win/Tie/Loss) for Ease Overall metric across all pairwise comparisons. Stacked bars show the proportion of comparisons won, tied, or lost by each platform.}
  \label{fig:preference_stack_ease_overall}
\end{figure}

\begin{figure}[htbp]
  \centering
  \includegraphics[width=0.8\linewidth]{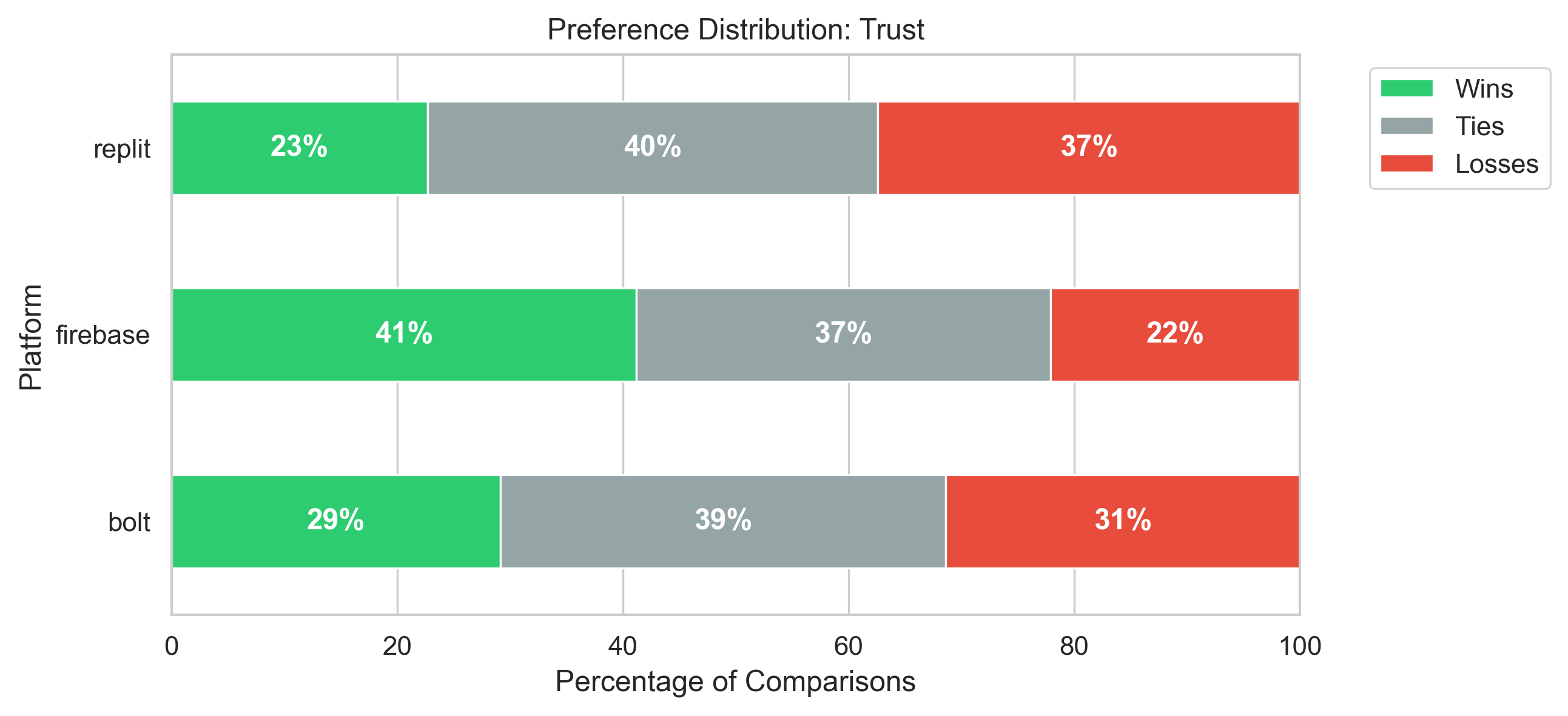}
  \caption{Distribution of comparison outcomes (Win/Tie/Loss) for Trust metric across all pairwise comparisons. Stacked bars show the proportion of comparisons won, tied, or lost by each platform.}
  \label{fig:preference_stack_trust}
\end{figure}

\begin{figure}[htbp]
  \centering
  \includegraphics[width=0.8\linewidth]{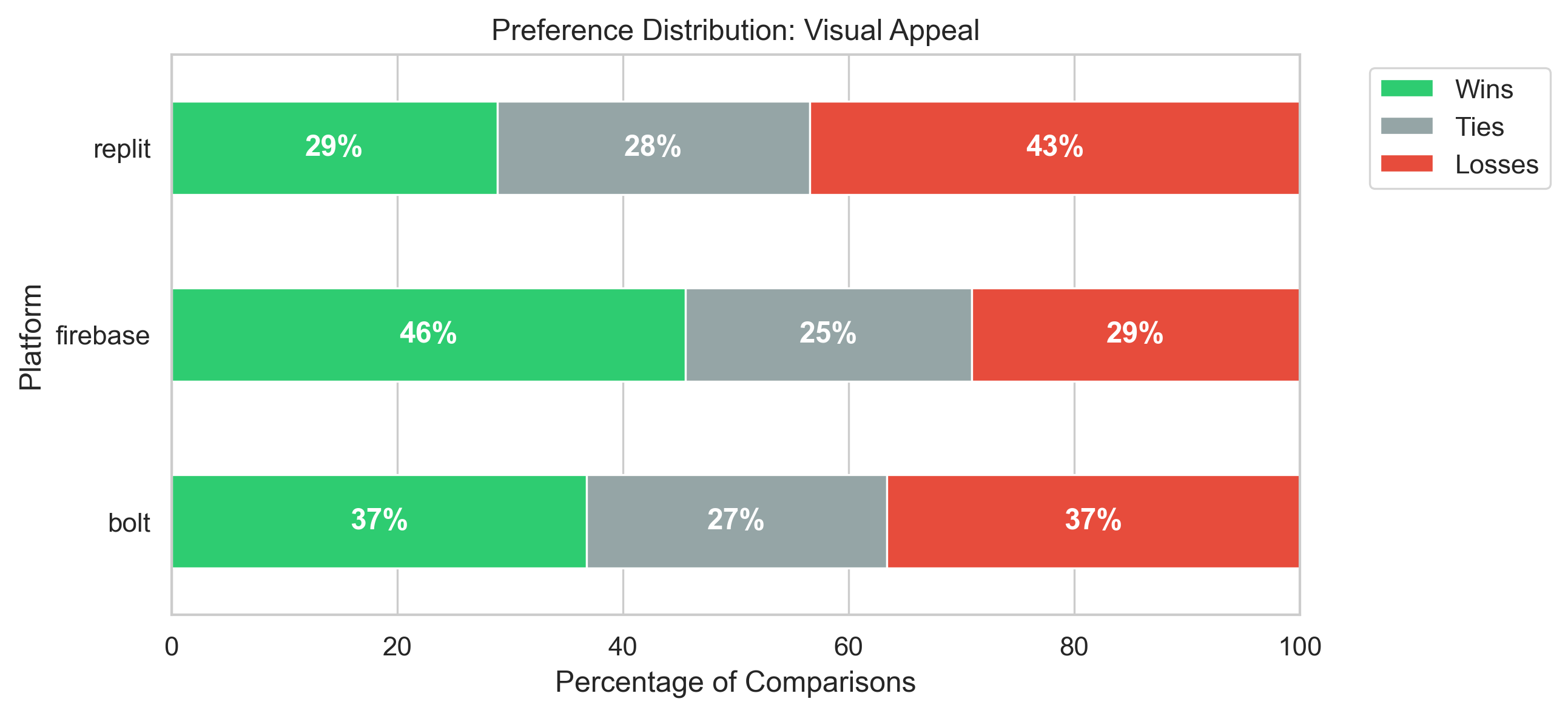}
  \caption{Distribution of comparison outcomes (Win/Tie/Loss) for Visual Appeal metric across all pairwise comparisons. Stacked bars show the proportion of comparisons won, tied, or lost by each platform.}
  \label{fig:preference_stack_visual_appeal}
\end{figure}

\begin{figure}[htbp]
  \centering
  \includegraphics[width=0.8\linewidth]{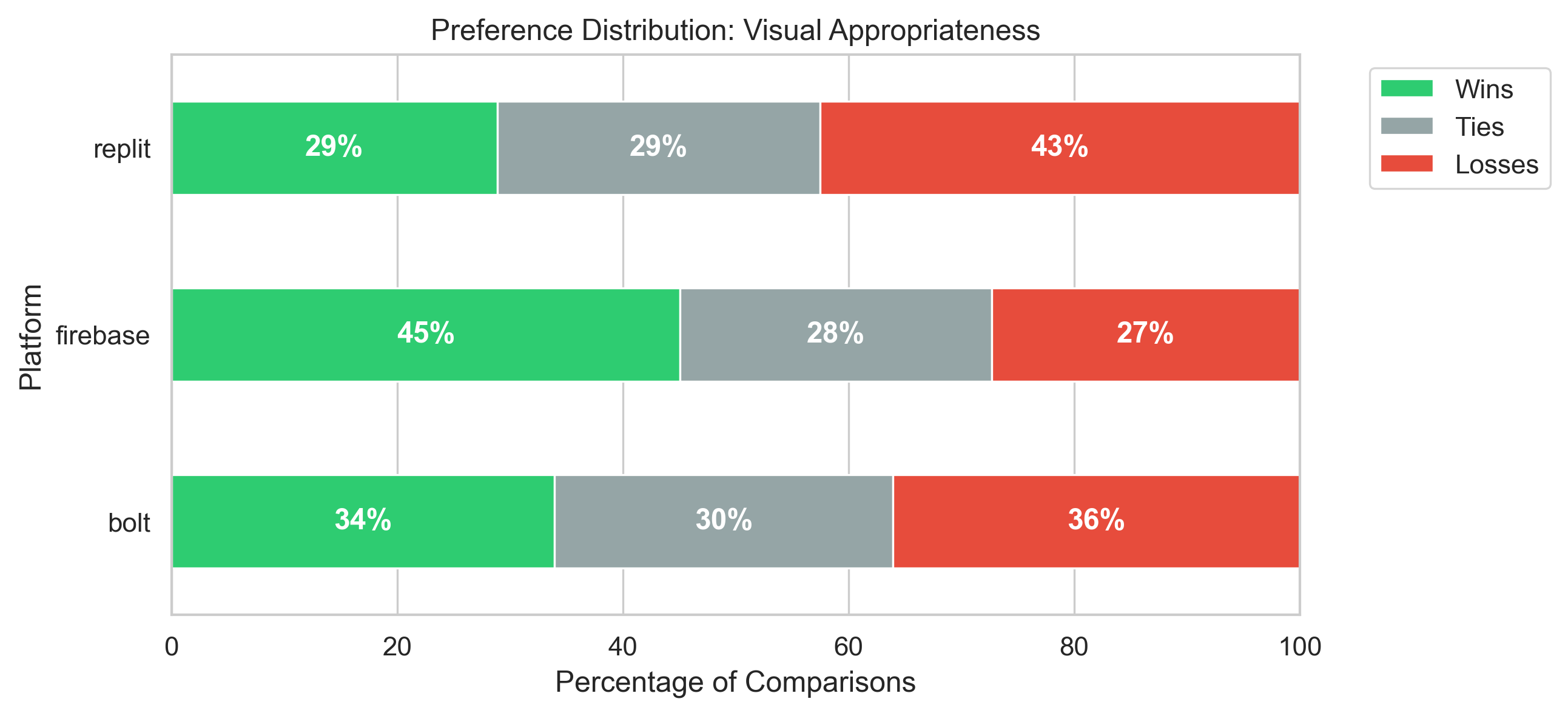}
  \caption{Distribution of comparison outcomes (Win/Tie/Loss) for Visual Appropriateness metric across all pairwise comparisons. Stacked bars show the proportion of comparisons won, tied, or lost by each platform.}
  \label{fig:preference_stack_visual_appropriateness}
\end{figure}

Figures \ref{fig:preference_stack_ease_overall}, \ref{fig:preference_stack_trust}, \ref{fig:preference_stack_visual_appeal}, and \ref{fig:preference_stack_visual_appropriateness} illustrate the preference distributions across Ease Overall, Trust, Visual Appeal, and Visual Appropriateness.

\subsubsection{Tournament Power Rankings: Trust and Visual Appeal}
\begin{figure}[htbp]
  \centering
  \includegraphics[width=\linewidth]{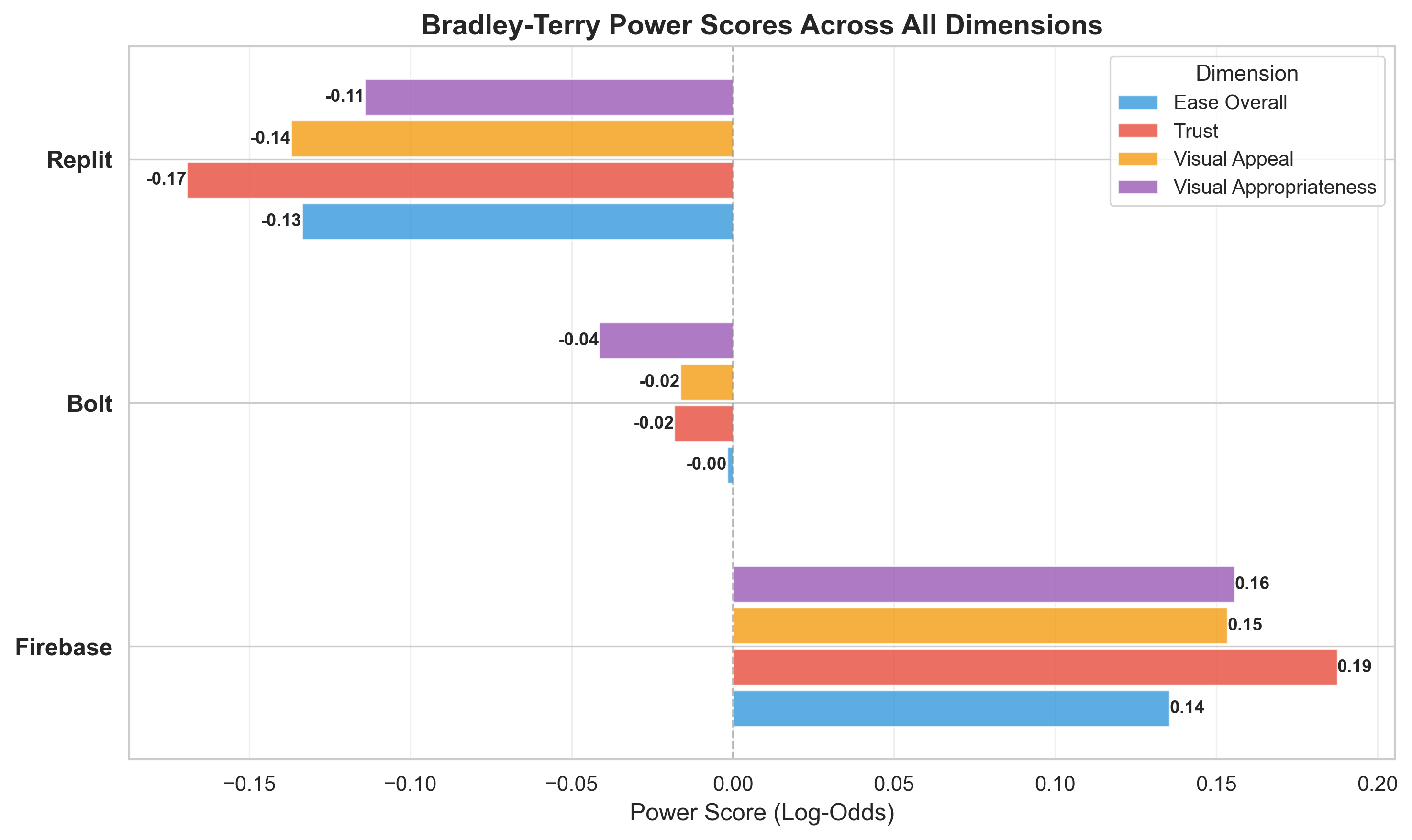}
  \caption{Bradley-Terry power scores across all dimensions (Ease Overall, Trust, Visual Appeal, Visual Appropriateness) for each platform. Power scores represent log-odds of winning comparisons. Higher scores indicate stronger performance in head-to-head matchups. Each platform has four bars, one for each dimension, color-coded by dimension.}
  \label{fig:bt_combined}
\end{figure}

To supplement the power rankings for overall ease and visual appropriateness, Figure \ref{fig:bt_combined} details the Bradley-Terry model results for user trust and subjective visual appeal. These figures confirm that Firebase maintains a statistically significant lead in the log-odds of winning head-to-head comparisons in both categories.

\subsubsection{Confidence Intervals and Detailed Win Rates}
\begin{figure}[htbp]
  \centering
  \includegraphics[width=\linewidth]{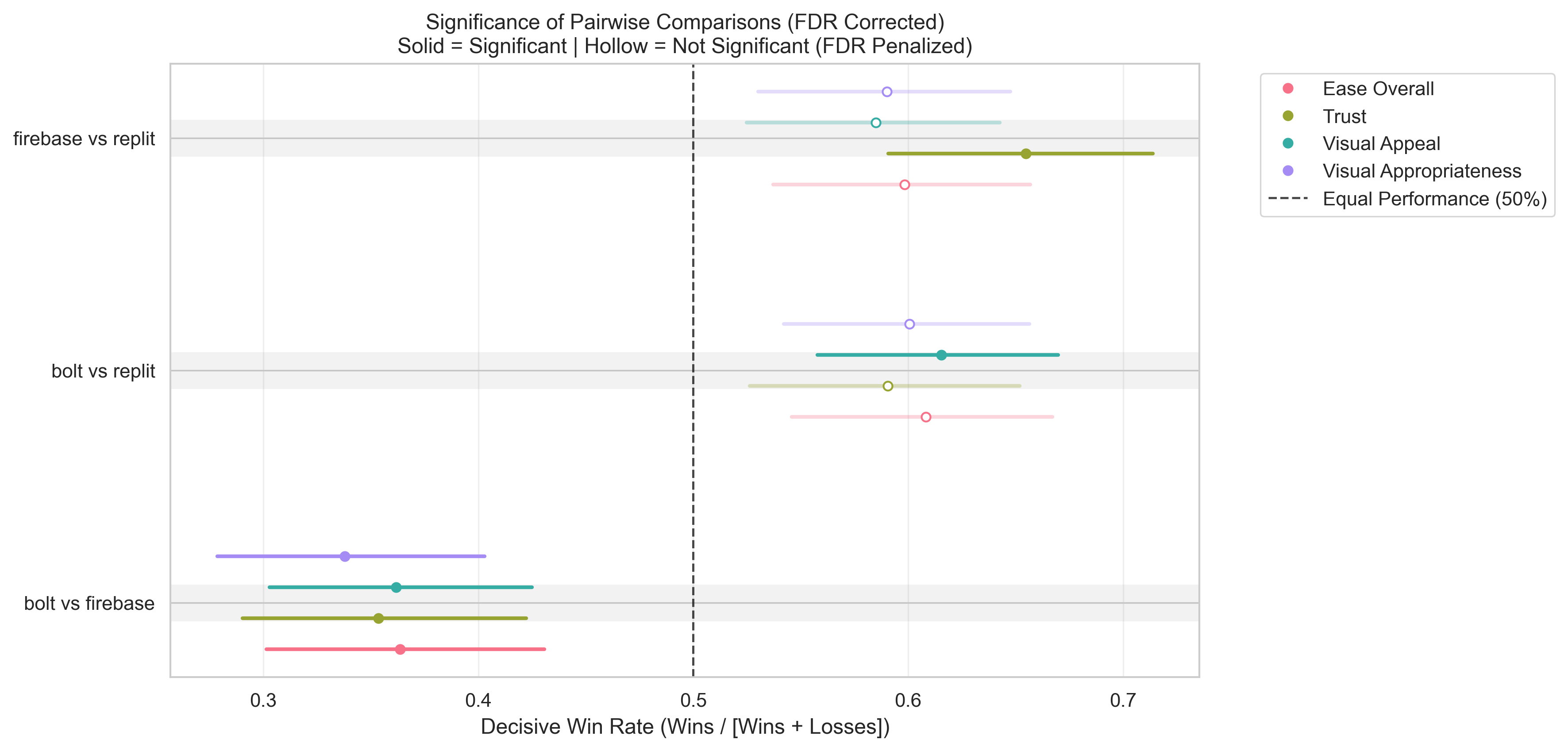}
  \caption{Forest plot of win rates with 95\% confidence intervals across all platform pairs. Estimates include only comparisons where both applications were successfully deployed.}
  \label{fig:forest_plot_win_rates_appendix}
\end{figure}

\begin{figure}[htbp]
  \centering
  \includegraphics[width=\linewidth]{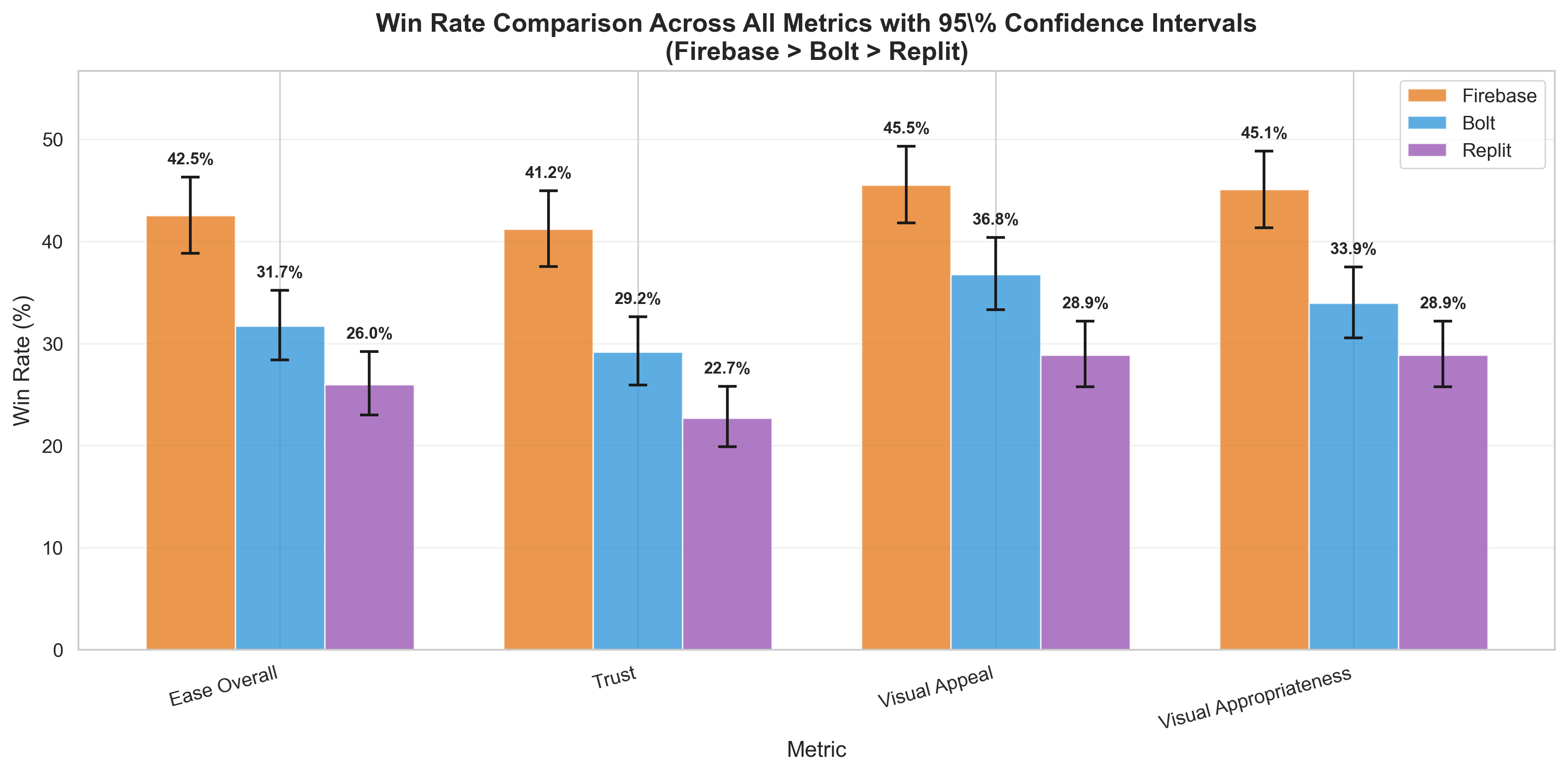}
  \caption{Comparison of win rates across all metrics with 95\% Wilson score confidence intervals, illustrating the comparative advantage of Firebase across Ease, Trust, and Aesthetics.}
  \label{fig:win_rate_comparison_with_ci_appendix}
\end{figure}

The following figures provide a more granular view of the win rates reported in Table \ref{tab:combined_win_rates_bt}. Figure \ref{fig:forest_plot_win_rates_appendix} presents a forest plot of the pairwise matchups, highlighting the consistency of Firebase's performance across different rater cohorts. Figure \ref{fig:win_rate_comparison_with_ci_appendix} presents the aggregate win rates for all platforms across all four primary comparison metrics, including 95\% Wilson score confidence intervals to denote statistical uncertainty.

\begin{table}[htbp]
    \centering
    \caption{Comprehensive platform performance summary combining Linear Mixed-Effects Model (LMM) absolute quality scores for Clarity and Ease metrics with average win rates across all comparison metrics. Overall rank is determined by composite score combining normalized LMM scores and win rates.}
    \label{tab:summary}
    \begin{tabular}{lcccc}
        \toprule
        Platform & LMM Clarity & LMM Ease & Avg Win Rate & Overall Rank \\
        \midrule
        Firebase & 3.92 & 3.68 & 43.6\% & 1 \\
        Bolt & 3.95 & 3.81 & 32.9\% & 2 \\
        Replit & 3.96 & 3.85 & 26.6\% & 3 \\
        \bottomrule
    \end{tabular}
\end{table}

\subsection{Leaderboard}

\begin{figure}[htbp]
  \centering
  \includegraphics[width=0.8\linewidth]{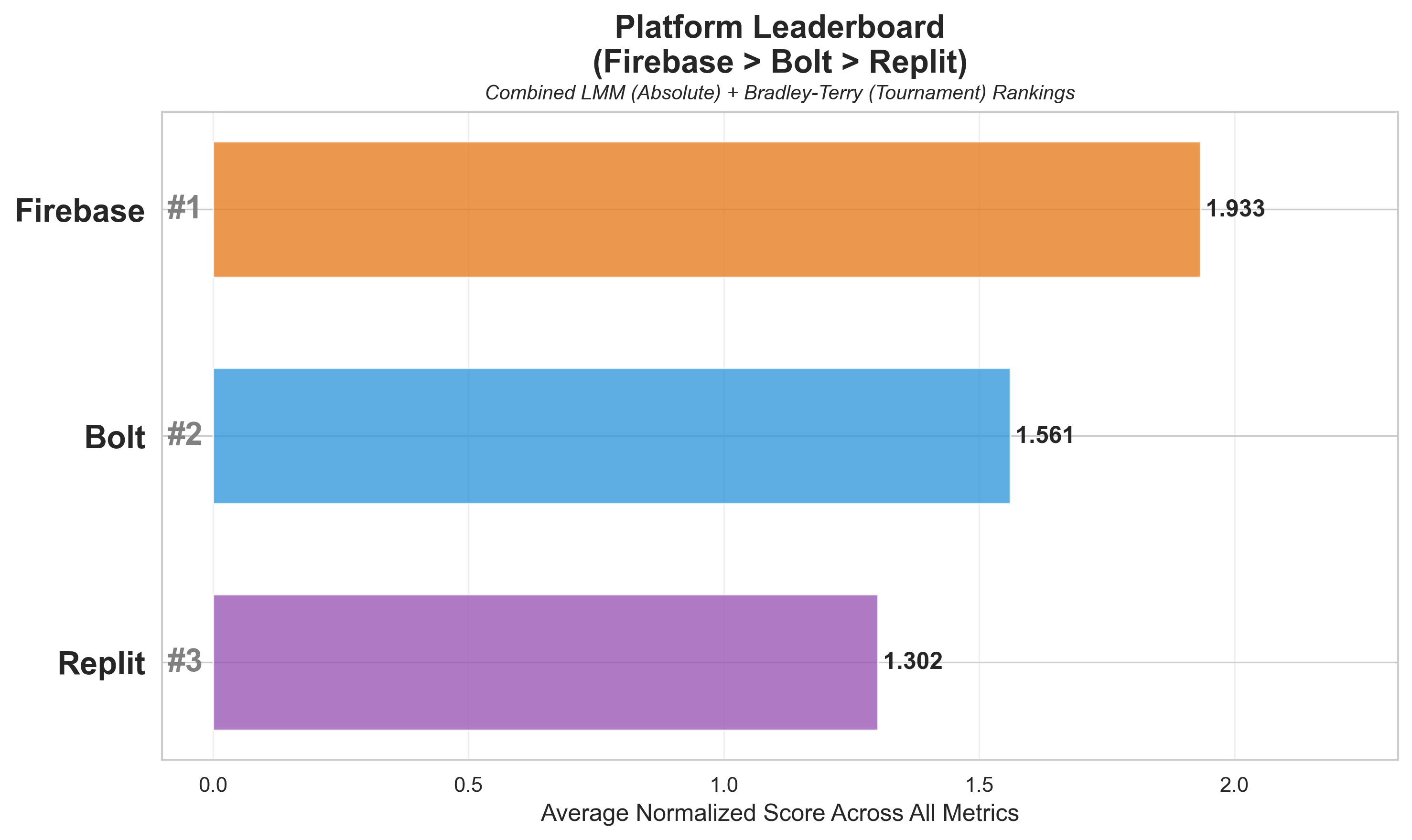}
  \caption{Overall platform ranking leaderboard combining normalized scores from Linear Mixed-Effects Models and Bradley-Terry tournament rankings across all metrics. Rankings show consistent Firebase $>$ Bolt $>$ Replit pattern across all statistical approaches.}
  \label{fig:platform_leaderboard}
\end{figure}

\paragraph{Leaderboard Culmination}
The final rankings, presented in Figure \ref{fig:platform_leaderboard} and Table \ref{tab:summary}, combine normalized LMM scores and Bradley-Terry tournament win rates. Firebase achieved the highest composite score (1.933), securing the \#1 position, followed by Bolt (1.561) and Replit (1.302). This result highlights Firebase's superior end-to-end coherence and its ability to translate visual clarity into a functional, trustworthy user experience.

\subsection{Prompt Dataset and Taxonomy Examples}
\label{sec:appendix_prompts}

This appendix provides a detailed overview of the prompt dataset used to evaluate the agentic ``prompt-to-app'' systems. To ensure a rigorous and diverse assessment, all prompts were categorized according to a multi-dimensional taxonomy involving Domain, Difficulty, Specificity, and Complexity. This structured approach allows for the evaluation of system performance across varied real-world scenarios, ranging from static informational sites to complex full-stack applications.

These prompts, all from the healthcare domain, are distributed across 8 unique combinations of the taxonomy dimensions to test specific edge cases, such as handling ambiguous requirements versus specific technical constraints. The following examples represent the specific prompts used to generate the artifacts in this analysis.

\begin{enumerate}

\item \textbf{Dimensions:} Difficulty=easy, Specificity=ambiguous, Complexity=high

\begin{quote}
We need a trustworthy and simple application for individuals managing ongoing medication schedules, such as those with chronic conditions or the elderly living independently. The goal is to create a ``digital pillbox'' that remembers their daily medications. This tool must be incredibly straightforward, requiring no technical skill to operate. The core idea is that a user, perhaps with a family member's help initially, can set up their list of medications once, and it will always be there when they open the app on their computer or tablet. This persistence is the key feature. The design should prioritize clarity and readability above all else. Think large fonts, clear separation of elements, and obvious controls. While it needs to save data, it should do so locally on the user's device to maintain privacy and avoid the complexity of accounts and logins. The user should feel a sense of security and control, knowing their personal medication list is stored only where they can access it. The specific layout and features for how they add, view, and manage this list are up to your interpretation.
\end{quote}

\item \textbf{Dimensions:} Difficulty=easy, Specificity=ambiguous, Complexity=low

\begin{quote}
We need a simple, privacy-focused tool for people who need temporary medication reminders. Imagine someone recovering from a minor procedure who only needs to take a specific antibiotic for one week, or an elderly person being cared for by a family member who just needs a quick way to set an alarm on a tablet for the next dose. The key here is ephemerality; the tool should not store any data whatsoever once the browser tab is closed. This is a single-session utility. Your goal is to create an application that feels as simple and disposable as a sticky note. It should be immediately obvious how to use it without any tutorial or explanation. The user should feel confident that their medical information isn't being saved anywhere, which is a major concern for this target demographic. The overall design should be calming, accessible, and instill a sense of reliability for its very limited, but important, purpose. How you choose to design the interface and implement the reminder mechanism is up to you, but the core principles of simplicity and zero data persistence are paramount.
\end{quote}

\item \textbf{Dimensions:} Difficulty=easy, Specificity=specific, Complexity=high

\begin{quote}
You are to build a persistent, client-side ``Daily Symptom Diary''. This application will allow a user to track their symptoms over time on their own device. The primary goal is to provide users with a simple tool to log how they are feeling each day and save this history for future reference, such as for an upcoming doctor's appointment. The application must save all entries to the browser's `localStorage`, ensuring that the user's diary is preserved between sessions. The interface should be straightforward, allowing users to easily add new entries and view their past logs in a clear, chronological order.

\#\#\#\# Core Functionality
* **New Entry Form**: The main page should feature a simple form to add a new symptom log.
    * A text input field for ``Symptom'' (e.g., ``Migraine'').
    * A number input or slider for ``Severity'' (on a scale of 1 to 10).
    * A `textarea` for optional ``Notes''.
    * An ``Add to Diary'' button.
* **Symptom History**:
    * Below the form, a section must display all past diary entries.
    * Each entry in the history must display the date it was logged, the symptom name, the severity, and the notes.
    * The entries must be sorted in reverse chronological order (most recent first).
    * A ``Delete'' button must be present for each entry, allowing the user to remove it from the diary.
* **Data Persistence**:
    * When the ``Add to Diary'' button is clicked, the new entry (including an automatic timestamp) is saved to an array in `localStorage`.
    * When the app loads, it must read the array from `localStorage` and render the complete symptom history.
    * Deleting an entry must also update the array in `localStorage`.

\#\#\#\# Data Model
* The diary will be stored in `localStorage` as a JSON array of objects.
* Each object will have four properties: `id` (a unique identifier, e.g., timestamp), `symptom` (string), `severity` (number), and `notes` (string).

\#\#\#\# Technical Constraints
* The application should be built with HTML, CSS, and vanilla JavaScript.
\end{quote}

\item \textbf{Dimensions:} Difficulty=easy, Specificity=specific, Complexity=low

\begin{quote}
Your task is to create a ``Single-Session Pill Reminder'', a simple, client-side web application. This tool is designed for users who need a quick, temporary reminder without wanting to create an account or store personal medical information. The core purpose is to set a one-time alert for a medication during their current browser session. All information should be cleared when the page is reloaded, ensuring complete privacy. This is crucial for users on public or shared computers. The application must have a clean, intuitive, and accessible interface that is easy for anyone, including the elderly or less tech-savvy individuals, to use without instructions. Think of it as a digital sticky note for medication.

\#\#\#\# Core Functionality
The application will have a single primary function: setting a one-time reminder.
* **Input Form**: The main interface will consist of a form with the following fields:
    * A text input field labeled ``Medication Name'' for the user to type the name of the pill (e.g., ``Ibuprofen'').
    * A time input field labeled ``Time'' for the user to select the time for the reminder.
    * A button labeled ``Set Reminder''.
* **Reminder Logic**:
    * When the ``Set Reminder'' button is clicked, the application should validate that both a name and a future time have been entered.
    * Upon successful validation, the app should display a confirmation message below the form, such as: ``Reminder set for Ibuprofen at 4:30 PM.''
    * At the exact time specified by the user, the browser must trigger a standard `alert()` dialog box. The alert should read: ``Time to take your [Medication Name]!''

\#\#\#\# UI/UX Requirements
* The layout should be a single, centered column.
* The design should be minimalist, with high-contrast text for readability.
* No navigation, additional pages, or complex settings are required.

\#\#\#\# Technical Constraints
* The application must be built using only HTML, CSS, and vanilla JavaScript.
* No external libraries or frameworks are allowed.
* The application must not use `localStorage`, `sessionStorage`, or cookies. All state must be held in memory for the current session only.
\end{quote}

\item \textbf{Dimensions:} Difficulty=hard, Specificity=ambiguous, Complexity=high

\begin{quote}
The challenge is to design and build a holistic, cloud-based platform for medication management and adherence. This system should serve as a vital link between patients, their families, and their healthcare providers. It needs to be a trustworthy, persistent, and intelligent system that goes beyond simple reminders. We envision a platform that not only reminds users to take their medication but also helps them understand their own habits and provides their care team with actionable insights. This requires a full-fledged application with user accounts, secure data storage, and a sophisticated feature set. Think about how to handle complex schedules, how to make the process of logging a dose as frictionless as possible, and how to present the resulting adherence data in a way that is meaningful for both a patient and their doctor. The architecture should be scalable and secure, as it will be handling sensitive personal health information. The specific design of the user interface, the choice of database, and the implementation of the features are in your hands, but the end product must be a comprehensive and reliable healthcare tool.
\end{quote}

\item \textbf{Dimensions:} Difficulty=hard, Specificity=ambiguous, Complexity=low

\begin{quote}
We need to build a powerful, single-session analytics tool for clinical pharmacologists and researchers. The purpose of this application is to help professionals understand complex medication adherence patterns without requiring any patient data to be stored. Imagine a researcher who has collected anonymized data from a study and needs a way to quickly visualize and explore it. The application should allow for the import of a dataset and then provide a suite of sophisticated visualization tools to uncover insights. The user should be able to explore the data from multiple angles, looking for correlations between non-adherence and factors like time of day or day of the week. The tool needs to feel professional and data-driven, providing clear, publication-quality graphics. However, it must function entirely within the browser session and guarantee that the sensitive (though anonymized) data is wiped clean when the session ends. The choice of specific charts, graphs, and analysis techniques is yours, but the focus must be on deep, insightful, and purely ephemeral data exploration.
\end{quote}

\item \textbf{Dimensions:} Difficulty=hard, Specificity=specific, Complexity=high

\begin{quote}
Build a comprehensive ``Chronic Condition Management Platform.'' This is a full-stack web application with a secure backend, designed for patients with chronic illnesses to track their symptoms, treatments, and lifestyle factors over the long term. The platform will require user accounts and will securely store all patient-entered data. Its core purpose is to provide a holistic view of the patient's health, enabling both the patient and their care team to identify triggers, track treatment efficacy, and make more informed health decisions. The system must support the logging of multiple data types, integrate with external APIs for environmental data, and provide a powerful analytics dashboard.

\#\#\#\# System Architecture
* **Frontend**: A responsive single-page application (SPA) using a framework like React or Vue.
* **Backend**: A secure RESTful or GraphQL API.
* **Database**: A time-series capable database (e.g., TimescaleDB, InfluxDB) or a standard relational database (e.g., PostgreSQL).

\#\#\#\# Core Features
* **User Authentication**: Secure signup/login system with encrypted passwords and session management.
* **Data Logging**:
    * A unified logging interface where users can add entries for:
        * Symptoms (e.g., pain level, fatigue).
        * Medications/Treatments (e.g., took 20mg of X).
        * Lifestyle Factors (e.g., diet, exercise, sleep quality).
    * All entries must be timestamped.
* **External API Integration**:
    * The system must automatically pull and store relevant environmental data based on the user's location (e.g., city), such as weather conditions and air quality index from a public API.
* **Analytics \& Correlation Dashboard**:
    * A dedicated section for data visualization.
    * Users must be able to generate charts that overlay their logged symptoms with other logged data (e.g., sleep quality vs. pain level).
    * The system should be able to run and display correlations between user-logged data and the automatically fetched environmental data.
* **Data Export**: Users must have the ability to export their complete data history as a CSV file.

\#\#\#\# Security \& Privacy
* All user data must be encrypted at rest and in transit.
* The application must be compliant with best practices for handling sensitive health data.
\end{quote}

\item \textbf{Dimensions:} Difficulty=hard, Specificity=specific, Complexity=low

\begin{quote}
Develop an advanced, single-session ``Medication Adherence Visualizer''. This is a sophisticated clinical tool for healthcare professionals to analyze a patient's medication schedule during a consultation. The app will not have a database or save any information; its purpose is to perform a complex, one-time analysis and visualization of an imported medication schedule. The professional will import a structured dataset (e.g., a CSV file) representing a patient's prescribed vs. actual intake times over a period. The application must then process this data and render several interactive charts and graphs to highlight patterns of adherence and non-adherence. The tool is designed for in-depth analysis within a single session, with all data being discarded upon refresh.

\#\#\#\# Core Functionality
* **Data Import**:
    * An ``Import Schedule'' button that opens a file selector.
    * The app must be able to parse a CSV file with the following columns: `medication\_name`, `prescribed\_time`, `actual\_intake\_time`.
* **Visualization Dashboard**: Once the data is imported, the app must render a dashboard with at least two of the following visualizations:
    * **Adherence Timeline Chart**: A chart that plots prescribed times vs. actual intake times over the period, showing deviations clearly.
    * **Time-of-Day Heatmap**: A heatmap showing at which hours of the day adherence is highest and lowest.
    * **Compliance Score Calculation**: A prominently displayed overall compliance percentage, calculated as the number of on-time doses divided by the total number of prescribed doses.
* **Interactivity**: The charts must be interactive. For instance, hovering over a data point on the timeline should show detailed information about that specific dose.

\#\#\#\# Technical Constraints
* Must use a charting library like D3.js, Chart.js, or a similar tool.
* The application must handle all data processing and visualization on the client side.
* No data should be sent to a server, and no information should be saved in the browser's storage. The tool is entirely stateless between sessions.
\end{quote}

\end{enumerate}

\subsection{Extended Participant Demographics}
\label{app:demographics}

\begin{figure}[htbp]
  \centering
  \begin{minipage}{0.48\linewidth}
    \centering
    \includegraphics[width=\linewidth]{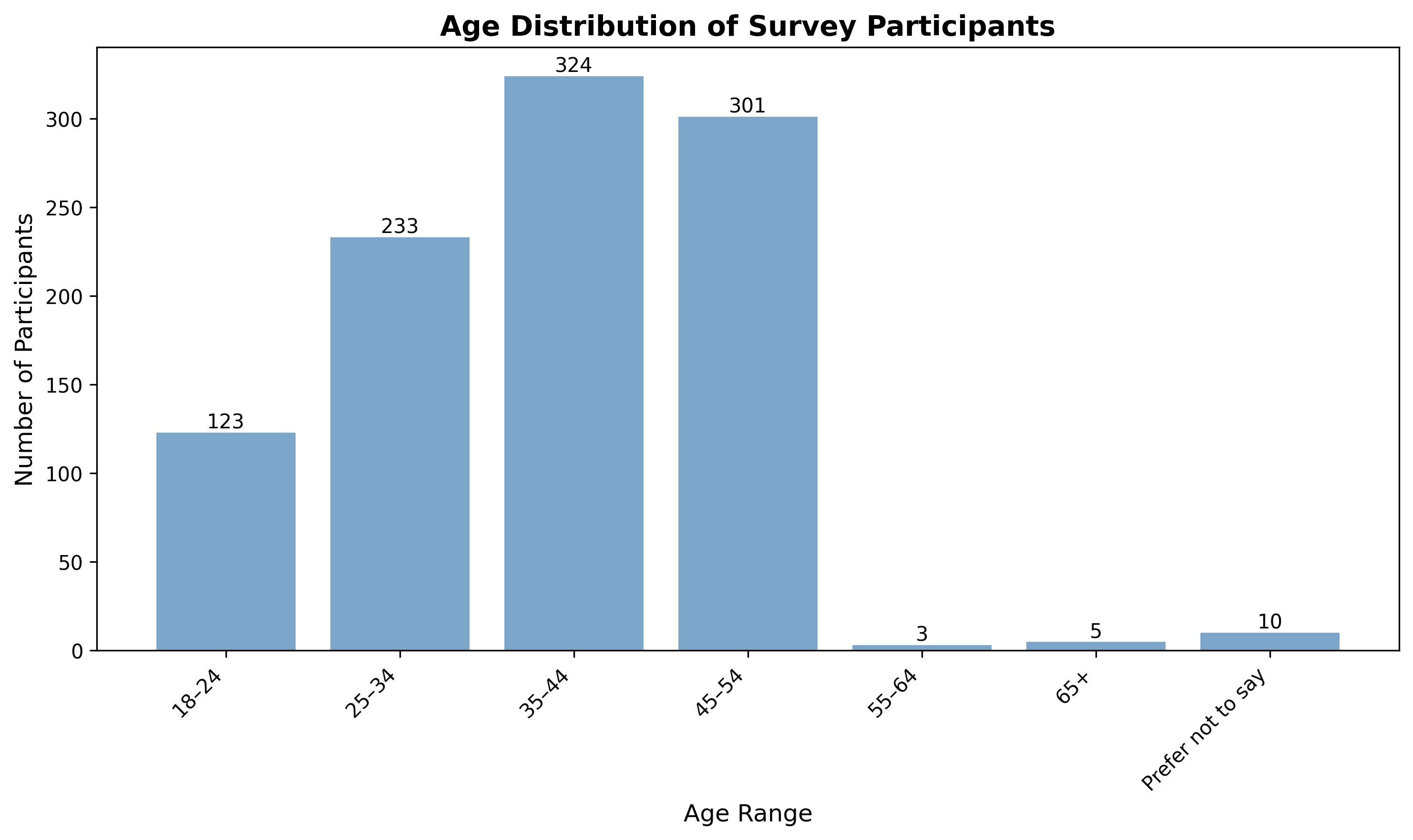}
  \end{minipage}\hfill
  \begin{minipage}{0.48\linewidth}
    \centering
    \includegraphics[width=\linewidth]{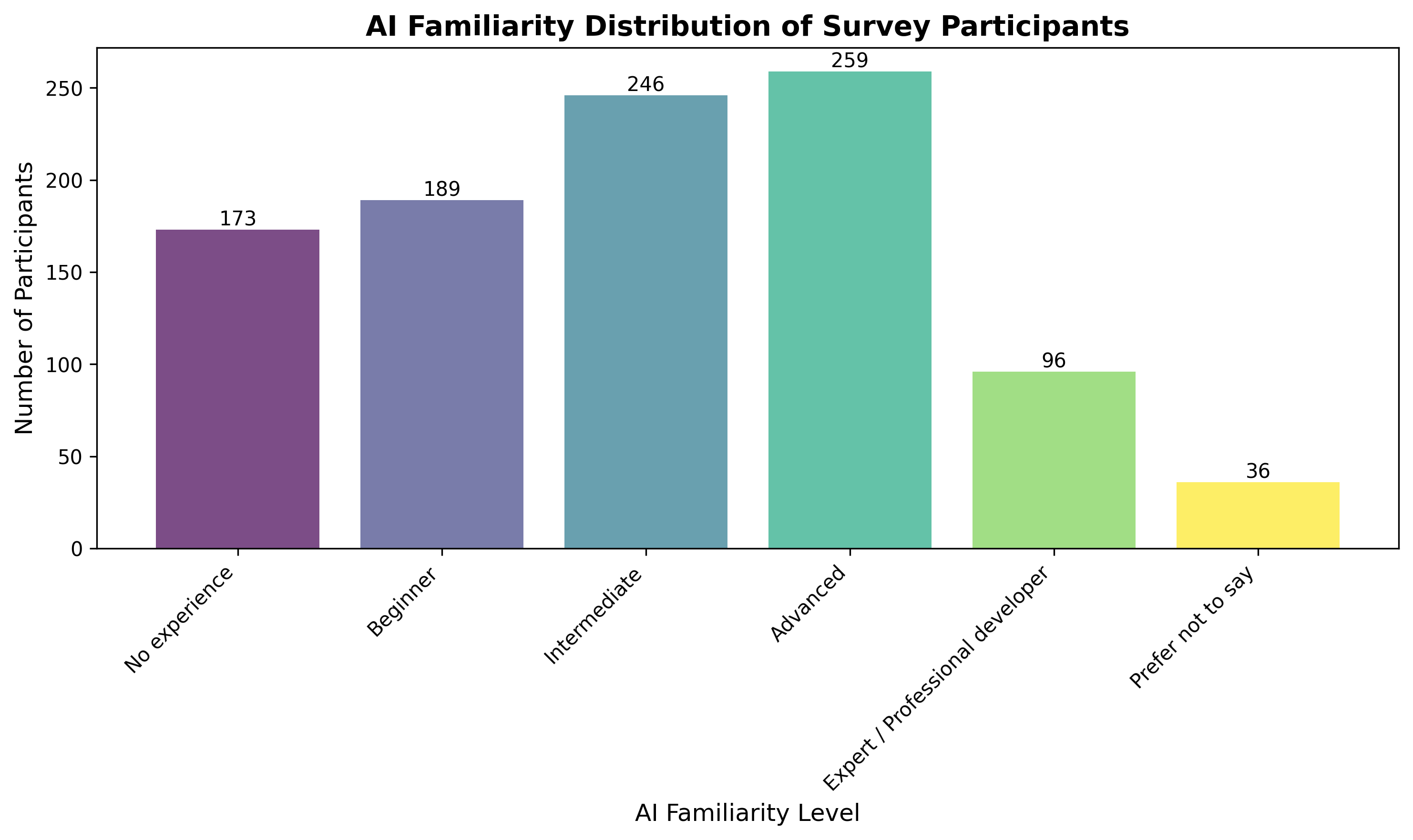}
  \end{minipage}
  \caption{Participant age distribution (left) and self-reported AI familiarity (right).}
  \label{fig:demographics_age_ai}
\end{figure}

\begin{figure}[htbp]
  \centering
  \begin{minipage}{0.48\linewidth}
    \centering
    \includegraphics[width=\linewidth]{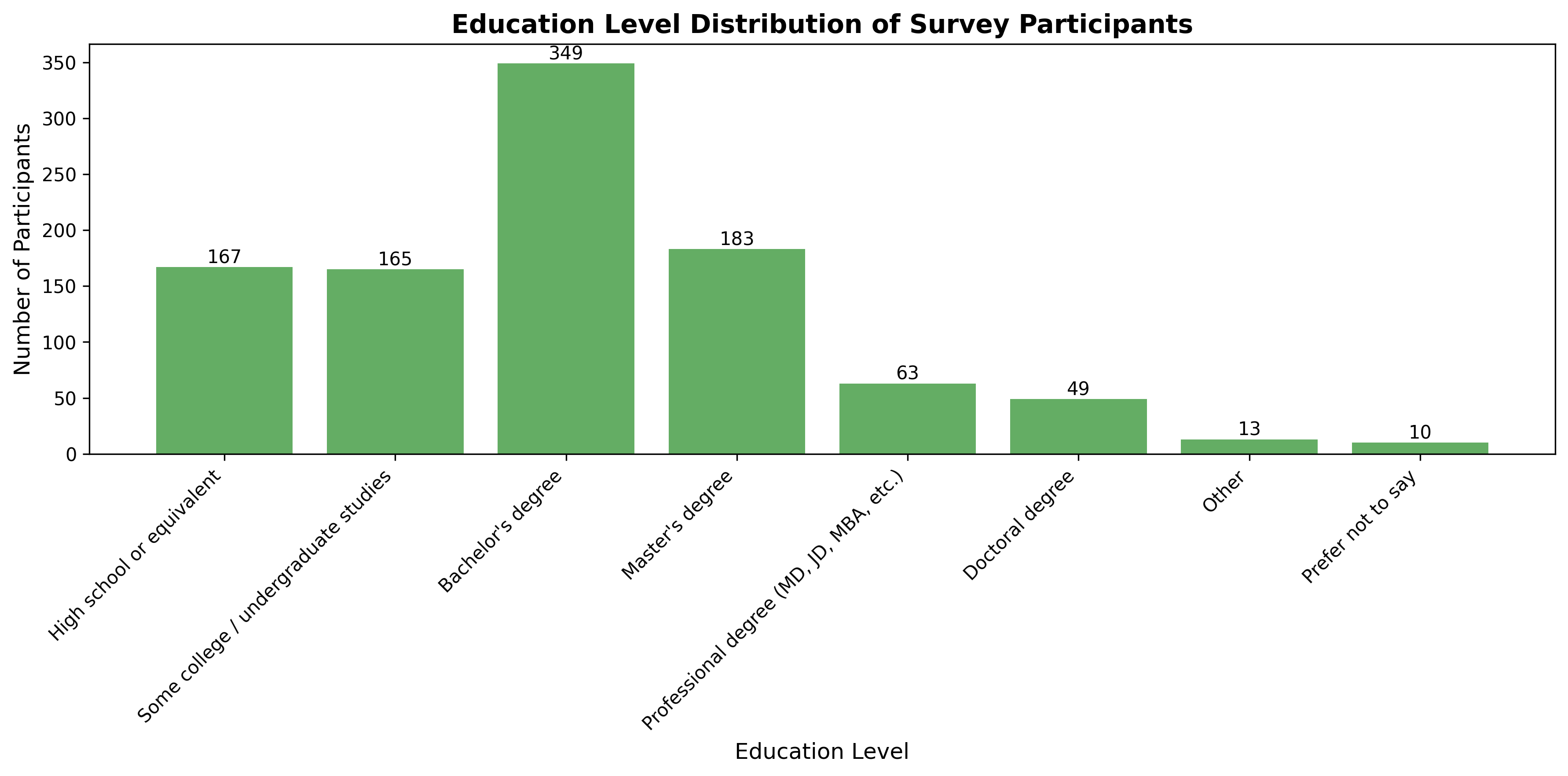}
  \end{minipage}\hfill
  \begin{minipage}{0.48\linewidth}
    \centering
    \includegraphics[width=\linewidth]{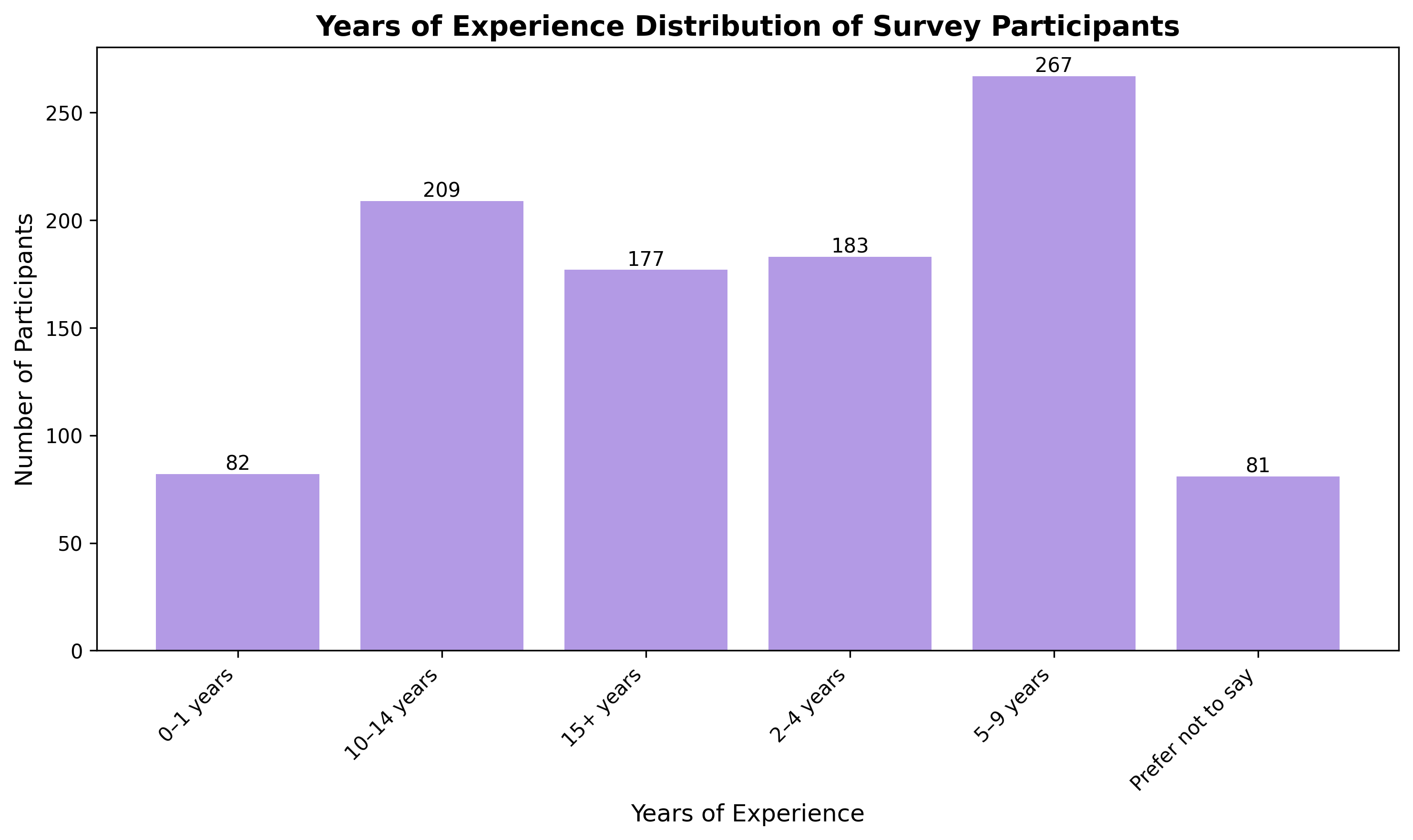}
  \end{minipage}
  \caption{Education level (left) and years of professional experience (right) of survey participants.}
  \label{fig:demographics_edu_exp}
\end{figure}

\begin{figure}[htbp]
  \centering
  \begin{minipage}{0.48\linewidth}
    \centering
    \includegraphics[width=\linewidth]{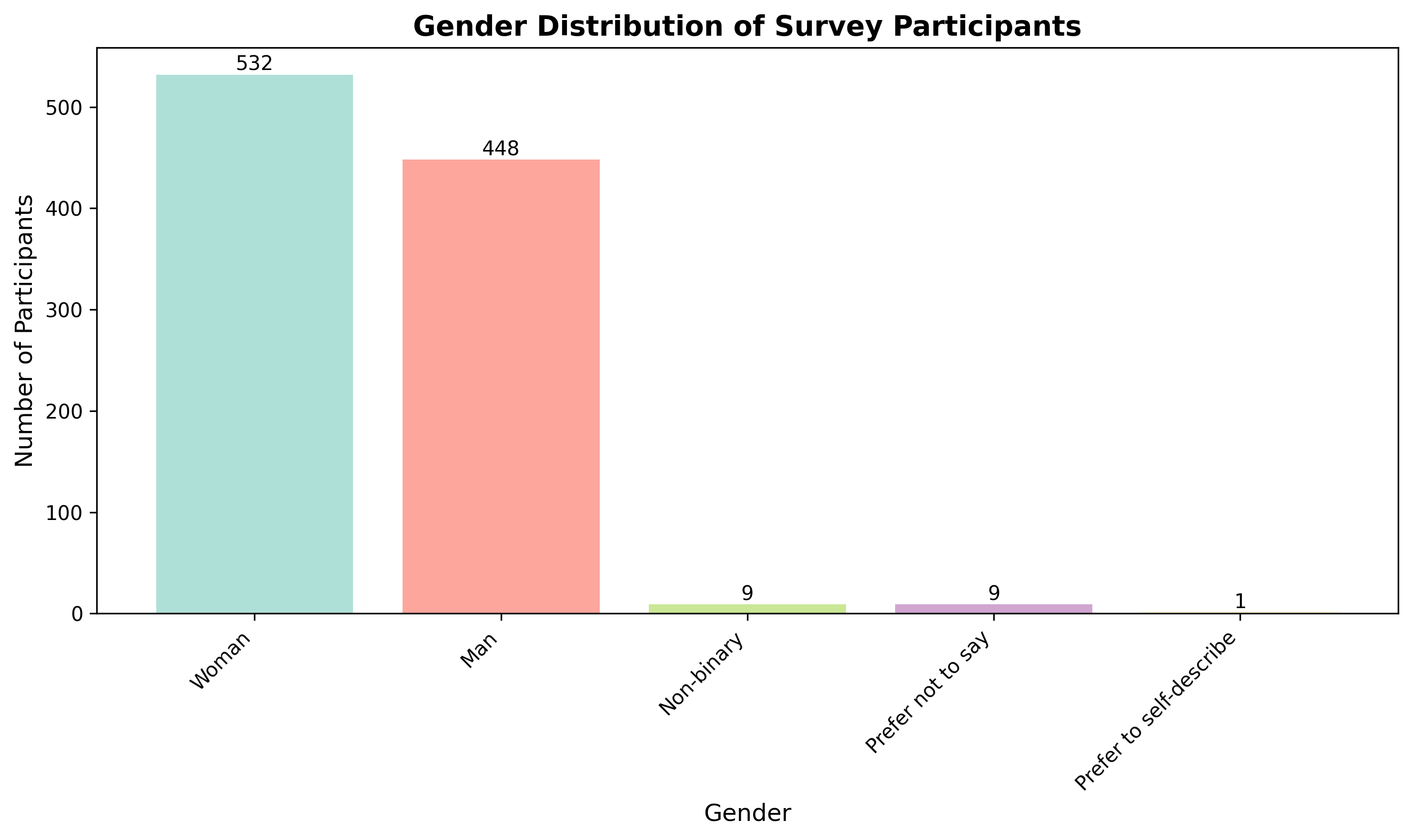}
  \end{minipage}\hfill
  \begin{minipage}{0.48\linewidth}
    \centering
    \includegraphics[width=\linewidth]{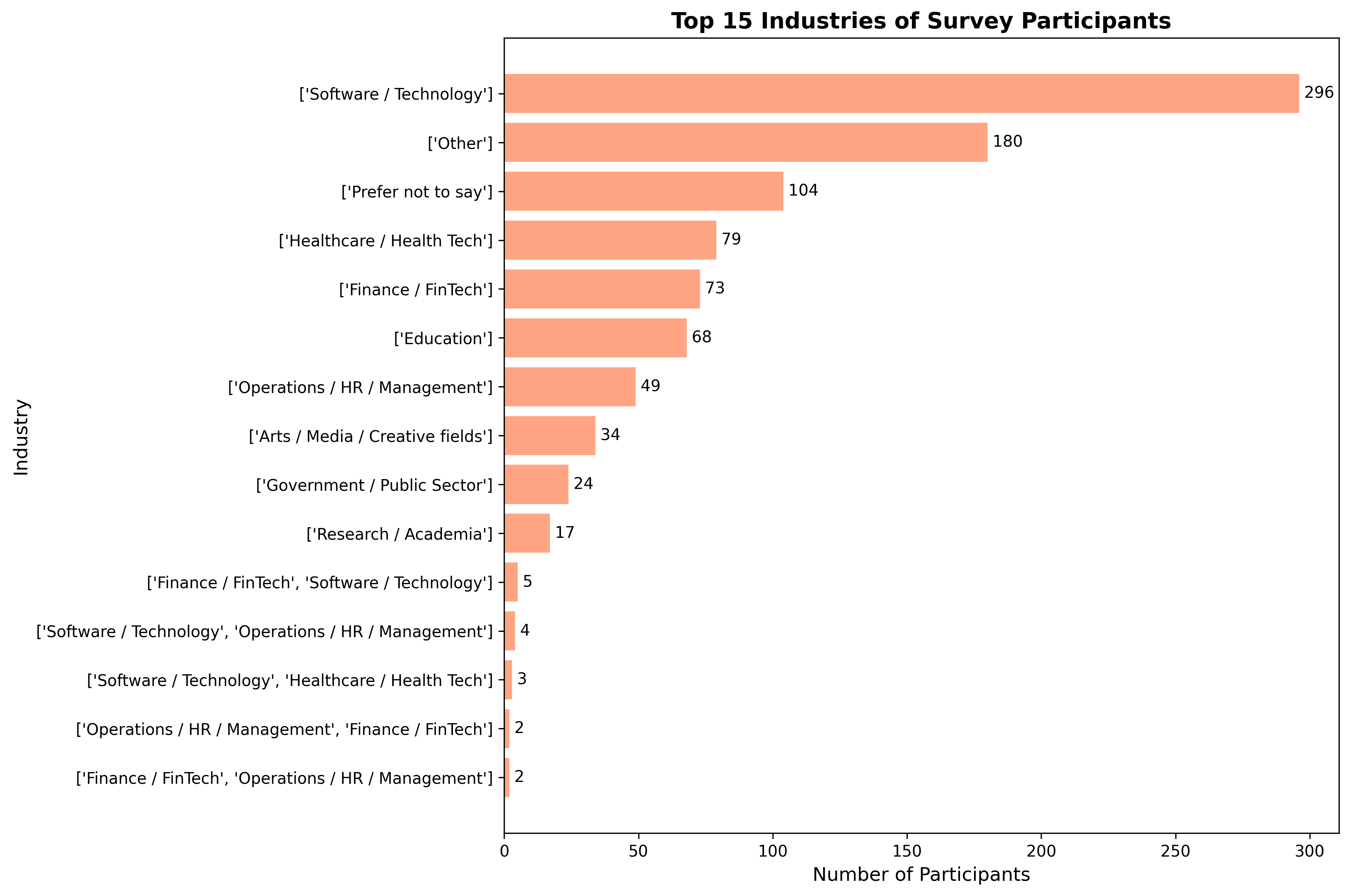}
  \end{minipage}
  \caption{Gender identity (left) and industry background (right) of survey participants.}
  \label{fig:demographics_gender_industry}
\end{figure}

This appendix provides extended demographic distributions for participants in the human-rater study, supplementing the summary shown in Figure~\ref{fig:demographic_overview} in the main text. 
Figures~\ref{fig:demographics_age_ai}--\ref{fig:demographics_gender_industry} report self-identified age ranges, gender, education level, industry background, years of professional experience, and familiarity with AI or software development tools.

These distributions are included to provide transparency regarding the composition of the rater pool. 
No subgroup or stratified analyses by demographic category were performed in this work, and the results reported in Section~\ref{sec:results} aggregate across all participants.

